\documentclass[%
 reprint,
superscriptaddress,
 amsmath,amssymb,
 aps,
pra,
floatfix,
longbibliography
]{revtex4-1}

\usepackage{ifpdf}
\usepackage{amsmath} 
\usepackage{amssymb}
\usepackage{amsfonts}
\usepackage{braket}
\usepackage{color}
\usepackage[colorlinks=true,linkcolor=blue,citecolor=blue,breaklinks]{hyperref}
\usepackage{graphicx}
\usepackage{dcolumn}
\usepackage{bm}
\usepackage[normalem]{ulem}

\newcommand{\la}{\langle}
\newcommand{\ra}{\rangle}
\newcommand{\tr}{\text{Tr}}

\newcommand{\TFD}{\mathbf{T}}

\definecolor{Zcolour}{rgb}{0.992, 0.588, 0.22}
\definecolor{purple}{rgb}{0.5,0,0.5}
\definecolor{brown}{rgb}{0.6,0.2,0}
\definecolor{dkgreen}{rgb}{0,0.5,0}

\usepackage{mathtools}

\begin{document}


\title{Entanglement renormalization of thermofield double states}
\author{Cheng-Ju Lin}
\affiliation{Perimeter Institute for Theoretical Physics, Waterloo, Ontario N2L 2Y5, Canada}
\author{Zhi Li}
\affiliation{Perimeter Institute for Theoretical Physics, Waterloo, Ontario N2L 2Y5, Canada}
\author{Timothy H. Hsieh}
\affiliation{Perimeter Institute for Theoretical Physics, Waterloo, Ontario N2L 2Y5, Canada}

\date{\today}

\begin{abstract}
Entanglement renormalization is a method for coarse-graining a quantum state in real space, with the multi-scale entanglement renormalization ansatz (MERA) as a notable example.  We obtain an entanglement renormalization scheme for finite-temperature (Gibbs) states by applying MERA to their canonical purification, the thermofield double state.  As an example, we find an analytically exact renormalization circuit for finite temperature two-dimensional toric code which maps it to a coarse-grained system with a renormalized higher temperature, thus explicitly demonstrating its lack of topological order.  Furthermore, we apply this scheme to one-dimensional free boson models at a finite temperature and find that the thermofield double corresponding to the critical thermal state is described by a Lifshitz theory. We numerically demonstrate the relevance and irrelevance of various perturbations under real space renormalization.
\end{abstract}

\maketitle

\textit{Introduction.---}Renormalization group (RG)~\cite{wilsonCritical1972} has proven to be an essential concept across many different disciplines of physics. 
In classical statistical mechanics, one RG scheme decimates some degrees of freedom~\cite{kadanoffScaling1966,cardyScaling1996} and yields an effective partition function on a coarse-grained system.
The idea of RG is also used to study the ground state and low-energy excitations of quantum systems, by using numerical renormalization group~\cite{wilsonRenormalization1975} or field theoretical methods~\cite{sachdevQuantum2011}, by decimating the high-energy degrees of freedom.

More recently, renormalization based on entanglement has proven to be both practically and conceptually useful.  For example, the density matrix renormalization group algorithm \cite{whiteDensity1992,whiteDensitymatrix1993} relies on entanglement considerations to decide which degrees of freedom are most relevant.     
Another major development is the multiscale entanglement renormalization ansatz (MERA)~\cite{vidalEntanglement2007,evenblyAlgorithms2009,evenblyEntanglement2009,evenblyClass2014}, which systematically constructs a quantum circuit to coarse grain a wavefunction in real space. 
MERA involves a hierarchy of local unitaries which remove short-range entanglement of the quantum state at different length scales, and can thus efficiently capture the  entanglement structure and other properties of  critical and gapped systems.
MERA (and the more general notion of entanglement renormalization (ER)) is not simply a numerical method.
The structure of MERA has appealing holographic interpretations~\cite{swingleEntanglement2012}, and ER has been applied to classify different quantum phases of matter at zero temperature~\cite{verstraeteRenormalizationgroup2005,chenLocal2010}, such as symmetry protected topological phases~\cite{chenClassification2011,schuchClassifying2011,singhSymmetryprotected2013,bridgemanAnomalies2017}, topological order~\cite{aguadoEntanglement2008,konigExact2009,liEntanglement2019} and fracton models~\cite{haahBifurcation2014,shirleyFracton2018,duaBifurcating2020}.

Despite the success of ER in studying ground state properties of quantum systems, its application to finite-temperature quantum systems is not as well-developed.
For such finite temperature quantum systems, one can write down a Landau-Ginzburg-Wilson theory which describes the vicinity of a phase transition and apply RG to the field theory.  However, as is the case for quantum ground states, it is valuable to have an RG approach in real space and based on entanglement considerations.   One would like an RG procedure which removes short-ranged entanglement from a thermal state, demonstrating in real space the relevance or irrelevance of various perturbations.

We obtain such a real-space, entanglement-based RG flow for thermal Gibbs states by considering the MERA for the canonical purification of the thermal state, namely the thermofield double (TFD) state. 
MERA provides an RG flow for the pure TFD state on successively coarse-grained lattices, and (by tracing out the auxiliary system) this provides a series of thermal density matrices on   coarse-grained systems.  

This procedure not only generates an RG flow for the thermal state but also provides an explicit circuit to construct a TFD state from a ``simple" fixed point state.  The TFD is an interesting state in itself, especially in the context of holography and wormholes~\cite{maldacenaEternal2003,maldacenaCool2013,lehnerGravitational2016,gaoTraversable2017,chapmanComplexity2017,chapmanComplexity2019a}, and our approach thus provides a complementary method to Refs.~\cite{wuVariational2019,cottrellHow2019a,martynProduct2019a,zhuGeneration2020a} for TFD preparation.
We will refer to this scheme as either ER of the thermal state or MERA of the TFD state.

We comment on the relations of our procedure to two previous pioneering works.
In Ref.~\cite{evenblyTensor2015}, Evenbly and Vidal applied tensor network renormalization to the two-dimensional (2d) classical partition function corresponding to a 1d finite temperature quantum system.  This indeed yields a MERA representation of the Gibbs state; however, such a MERA was not designed to provide an RG flow for the Gibbs state.  In particular, the unitaries in their MERA do not alter the spectrum and entropy of the thermal state.  In contrast, as we will show, our MERA on the thermofield double state does enable the thermal spectrum to change.  
In Refs.~\cite{swingleRenormalization2016a,swingleMixed2016}, Swingle and McGreevy pioneered the ``s-sourcery" framework which characterizes the complexity of a pure or mixed quantum state. In their mixed s-sourcery formalism, one definition considers the unitary circuit preparing the purification of the mixed state. 
Our proposal is therefore an explicit construction of such a circuit based on MERA.
The RG circuits constructed in Ref.~\cite{PhysRevB.93.205159} for some classical statistical models can also be generalized to the TFD setting as discussed in Ref.~\cite{swingleMixed2016}.

In particular, we apply ER to two nontrivial thermal systems, showcasing its potential.
First, we construct an analytic, {\bf exact} ER circuit which maps the finite temperature 2d toric-code state~\cite{KITAEV20032} onto a coarse-grained lattice with a renormalized temperature.  Our exact construction explicitly   
reveals the RG flow of the finite-temperature toric code to infinite temperature. As a second example, we apply ER to a finite temperature free boson model in 1d and find that the TFD corresponding to the critical thermal state is described by a Lifshitz theory. 
We find that the ER procedure is consistent with momentum space RG and in particular demonstrates the relevance and irrelevance of various perturbations. By removing short-range entanglement, the ER reveals the RG flow to a purely classical model at long length scales. 

\begin{figure}
    \includegraphics[width=\columnwidth]{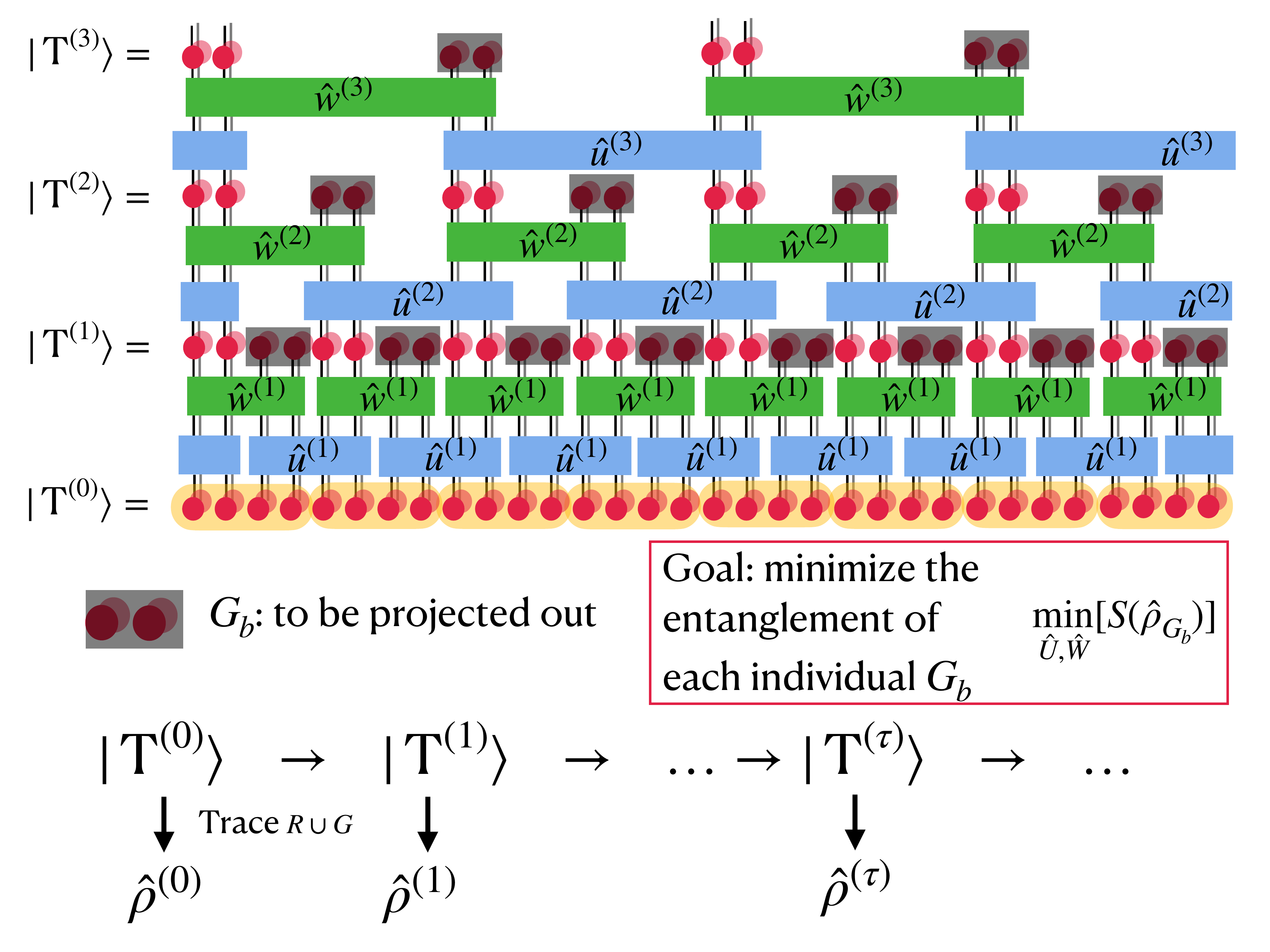}
    \caption{{\bf ER of Thermal State from MERA of Thermofield Double State} A one-dimensional illustration of the structure of the MERA circuit on the TFD state. The circuit is composed of ``disentanglers" $\hat{U}$ and ``isometries" $\hat{W}$, chosen so that individual ``garbage" blocks $G$ are minimally entangled with the rest of the system.  By tracing out the auxiliary system $R$, one obtains a sequence of thermal states on successively coarse-grained lattices. 
    }    \label{fig:ER}
\end{figure}

\textit{Setup.---} Given a Gibbs ensemble $\hat{\rho} = \frac{1}{Z_\beta} e^{-\beta \hat{H}}$, where $Z_\beta=\tr[e^{-\beta \hat{H}}]$, we consider the canonical purification or the thermofield double state:
\begin{align}\label{eqn:generalTFD}
    |\TFD \ra &\equiv \frac{1}{\sqrt{Z_{\beta}}}\sum_{E}e^{-\frac{\beta}{2} E}|E\ra_L |E\ra_R 
    = \sqrt{\frac{Z_{\beta=0}}{Z_{\beta}}}e^{-\frac{\beta}{2}\hat{H}_L}|\TFD_{\beta=0}\ra~,
\end{align}
where $|E\ra_{L,R}$ are eigenstates of $\hat{H}$ with energy $E$ on the original (L) and identical auxiliary (R) systems, respectively. $|\TFD_{\beta=0}\ra~$ is a maximally entangled state between $L$ and $R$; for example, in a qubit system, one choice is $|\TFD_{\beta=0}\ra=\frac{1}{\sqrt{Z_{\beta=0}}}\sum_{\boldsymbol{\sigma}}|\boldsymbol{\sigma} \ra_L | \boldsymbol{\sigma} \ra_R$ for the configurations $\boldsymbol{\sigma} = (\sigma_1, ... ,\sigma_N)$, where $\sigma_i= \downarrow$ or $\uparrow$ is the state for $i$-th qubit.  The TFD is a purification of the Gibbs state because $\hat{\rho}=\tr_R |\TFD\ra\la \TFD| $.

We consider a MERA circuit for disentangling the TFD state, illustrated in Fig.~\ref{fig:ER} for 1d. 
The TFD state is supported on a lattice in which each site consists of $L$ and $R$ degrees of freedom. 
We group the lattice sites into blocks of $2M$ sites, with block index $b = 1 \dots N/(2M)$.
Each layer of the circuit consists of ``disentanglers" $\hat{U}=\bigotimes_b \hat{u}_{b,b+1} $, where $\hat{u}_{b,b+1}$ is a unitary gate operating across the boundaries of the blocks, and ``isometries" $\hat{W}=\bigotimes_b \hat{w}_{b}$, where $\hat{w}_{b}$ is an unitary gate operating within the block.  The image of each $\hat{w}_b$ is divided into degrees of freedom which are decoupled (denoted $G_b$) and those that remain (see Fig.~\ref{fig:ER}).
Here, we consider the gates which preserve the $L \leftrightarrow R$ swap symmetry of the TFD. 
Unlike in most ground state algorithms, where MERA is obtained by minimizing the energy, here the circuit is obtained by minimizing the entanglement entropy of each individual ``garbage" block $G_b$ from the rest of the system as much as possible.
That is, for each layer, we search for $\hat{U}$ and $\hat{W}$ operating on the TFD ($|\TFD^{\prime}\ra = \hat{W}\hat{U}|\TFD\ra$) to minimize $S(\rho_{G_b})= -\tr (\hat{\rho_{G_b}}\ln (\hat{\rho_{G_b}}))$, the entanglement entropy of the ``garbage" block  $\hat{\rho}_{G_b} \equiv \tr_{G_b^c}[|\TFD^{\prime}\ra \la \TFD^{\prime}|]$.

Iterating this procedure yields a sequence of states $|\TFD^{(0)}\ra \rightarrow |\TFD^{(1)}\ra \dots \rightarrow |\TFD^{(\tau)}\ra...$ on successively coarse-grained lattices, and this in turn yields a sequence of mixed states $\hat{\rho}^{(0)} \rightarrow \hat{\rho}^{(1)} \dots \rightarrow \hat{\rho}^{(\tau)}...$ obtained by tracing out $R \cup G$ where $G=\bigcup_b G_b$ (e.g. $\hat{\rho}'= \tr_{R\cup G}[|\TFD^{\prime} \ra \la\TFD^{\prime}|]$).  Because each iteration involves unitaries which are local, long-distance properties are maintained after every step.  Hence, the above sequence of mixed states is the desired RG flow of the thermal state.  By rewriting $\hat{\rho}^{\prime}=\frac{1}{Z^{\prime}}e^{-\beta^{\prime} \hat{H}^{\prime}}$, one can also obtain the renormalized temperature $\beta^{\prime}$ and renormalized Hamiltonian $\hat{H}^{\prime}$, and obtain the RG flow. 
While we illustrate the procedure in 1d above and in Fig.~\ref{fig:ER}, it is easy to generalize it to higher dimensions~\cite{evenblyAlgorithms2009,evenblyEntanglement2009}.

Note that our setup generally involves unitaries acting between $L$ and $R$.
The entanglement spectrum across $L$ and $R$ is in fact the thermodynamic spectrum of the reduced density matrix. 
The gates operating between $L$ and $R$ therefore allow the thermodynamic spectrum to change, generating a nontrivial RG flow.
Without such mixing, the entanglement between $L$ and $R$, which is the thermodynamic entropy of the Gibbs state, is fixed. In that case, the initial entropy would be concentrated on successively fewer degrees of freedom, leading to infinite temperature in all cases.

The techniques used in MERA can be straightforwardly applied to this procedure, and therefore any local observables in the coarse-grained system can be efficiently calculated.


\begin{figure}
    \includegraphics[width=\columnwidth]{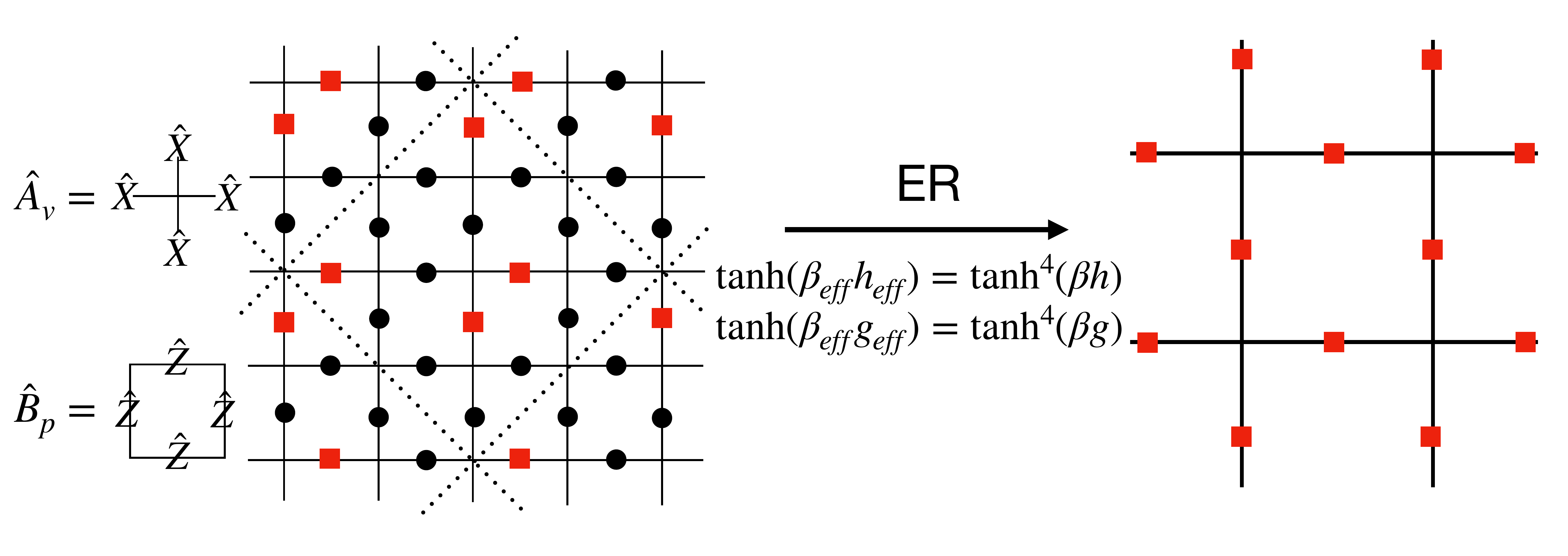}
    \caption{{\bf ER of Finite Temperature 2d Toric Code} We find an exact ER circuit (Appendix~\ref{app:disentagle_circuit}) which produces the same model on a coarse-grained lattice with renormalized temperature specified above. } 
    \label{fig:toric}
\end{figure}

\textit{ER of Toric code Gibbs state.---}Our first application is an exact ER of the toric code model~\cite{KITAEV20032} at a finite temperature. 
Consider a square lattice in 2d where the qubit degrees of freedom reside on the links.
The toric code Hamiltonian is defined as:
\begin{equation}\label{eqn:toric_code}
    \hat{H}=-h\sum_v \hat{A}_v - g\sum_p \hat{B}_p~,
\end{equation}
where $\hat{A}_v$ and $\hat{B}_p$ are the ``star" and ``plaquette" terms defined as in Fig~\ref{fig:toric}.  The topologically ordered ground states of toric code can be exactly coarse-grained by an MERA circuit as shown in Ref.~\cite{aguadoEntanglement2008}. 

Remarkably, we are able to exactly disentangle the TFD state of the finite-temperature toric code, defined in Eq.~(\ref{eqn:generalTFD}), with $\hat{H}_L$ being the toric code Hamiltonian Eq.~(\ref{eqn:toric_code}) operating on the $L$ side.
At each RG step, we aim to ``decimate" the black lattice sites as shown in Fig~\ref{fig:toric}.  As a reminder, each site of the doubled system consists of two qubits.
The first part of our circuit is constructed from gates acting within $L$ and $R$ independently and is in fact the same circuit (up to some swaps of the qubits) given in Ref.~\cite{aguadoEntanglement2008} to disentangle the toric code ground state. 
However, to disentangle the TFD at finite temperature, one further requires gates operating between $L$ and $R$ sides.  The full circuit is very technically involved, and we present all the details explicitly in the Appendix~\ref{app:disentagle_circuit}.

Our exact RG circuit generates a mapping of the Gibbs ensemble to a coarse grained lattice with the same form of the Hamiltonian [Eq.~(\ref{eqn:toric_code})] but with renormalized effective temperature and couplings:
\begin{equation}\label{exactRGE}
    \tanh(\beta_{\text{eff}}h_{\text{eff}}) = \tanh^{4}(\beta h)~,~    \tanh(\beta_{\text{eff}}g_{\text{eff}}) = \tanh^{4}(\beta g)~.
\end{equation}
[We find that the star and plaquette terms renormalize independently. To generate the effective star (plaquette) terms, only the star (plaquette) terms are involved.]
This exact ER shows explicitly that the finite-temperature 2d toric code flows to the infinite-temperature ensemble, providing another way of showing that it is topologically trivial, complementing the results of Ref.~\cite{hastingsTopological2011, luDetecting2020,PhysRevB.76.184442}.

While a topologically trivial TFD state implies a topologically trivial thermal ensemble (in Ref.~\cite{hastingsTopological2011}'s definition), the converse does not hold. Here we give an explicit example to show that a topologically trivial ensemble, albeit with classical order, can have a topologically nontrivial TFD.

The example we consider here is again the toric code model Eq.~(\ref{eqn:toric_code}), but with $h\!=\!\infty$ and at infinite temperature.  Specifically, we take the limit $\beta h=\infty, \beta g=0$.  This thermal state is maximally mixed within the subspace of states satisfying $\hat{A}_v |\psi \ra =|\psi \ra$.
This state has classical long-range order, as witnessed by correlation functions of $\hat{X}$-type string operators. (Alternatively, it is a deconfined classical gauge theory \cite{wegnerDuality1971}.) However, the thermofield double state for this system is:
\begin{equation}
    |\TFD \ra\propto\sum_{\boldsymbol{\sigma}|\hat{A}_v=1}|\boldsymbol{\sigma}  \ra_{L}|\boldsymbol{\sigma}  \ra_{R},
\end{equation}
where $\boldsymbol{\sigma}$ runs over spin configurations in $\hat{X}$ basis such that each $\hat{A}_v=1$.

As this is a coherent superposition of loop configurations, we expect it to have topological order.  Indeed, consider the state $|\psi\ra\propto\sum_{\boldsymbol{\sigma}|\hat{A}_v=1}|\boldsymbol{\sigma}  \ra$, which is a ground state of the full toric code model Eq.~(\ref{eqn:toric_code}) with $h,g>0$. $|\psi \ra$ is manifestly topologically ordered, as revealed by its nonzero topological entanglement entropy, for example ~\cite{PhysRevLett.96.110404,PhysRevLett.96.110405}. For any calculation in the $\hat{X}$ basis, $|\TFD \ra$ and $|\psi\ra$ are essentially the same. For example, the reduced density matrix of $|\psi\ra$ in a subregion $\mathcal{A}$ is:
\begin{equation}
    \hat{\rho}_{\mathcal{A}}\propto\sum_{\boldsymbol{\sigma}_1\boldsymbol{\sigma'}_1\boldsymbol{\sigma}_2}|\boldsymbol{\sigma}_1  \ra\la\boldsymbol{\sigma'}_1|,
\end{equation}
where $\boldsymbol{\sigma}_1$ and $\boldsymbol{\sigma'}_1$ are spin configurations inside ${\mathcal{A}}$, $\boldsymbol{\sigma}_2$ are spin configurations outside, and the summation runs over spin configurations such that both $\boldsymbol{\sigma}_1\boldsymbol{\sigma}_2$ and $\boldsymbol{\sigma'}_1\boldsymbol{\sigma}_2$ satisfy $\hat{A}_v=1$. Similarly, the reduced density matrix of $ |\TFD \ra$ is given by formally rewriting $\boldsymbol{\sigma}$ as $\boldsymbol{\sigma}_L\boldsymbol{\sigma}_R$. Therefore, all entanglement entropies and especially the topological entanglement entropy for $|\TFD \ra$ and $|\psi \ra$ are the same.

We conclude that the thermofield double state has topological order even though the thermal subsystem has only classical order.
From the exact RG equations Eqs.~(\ref{exactRGE}), we can also conclude that $(h=\infty, g>0, \beta>0)$ will be in the same phase.
We note that 3d toric code at a finite temperature exhibits classical order below a critical temperature, and we expect our results above to also apply in this case.

\begin{figure*}
    \includegraphics[width=2\columnwidth]{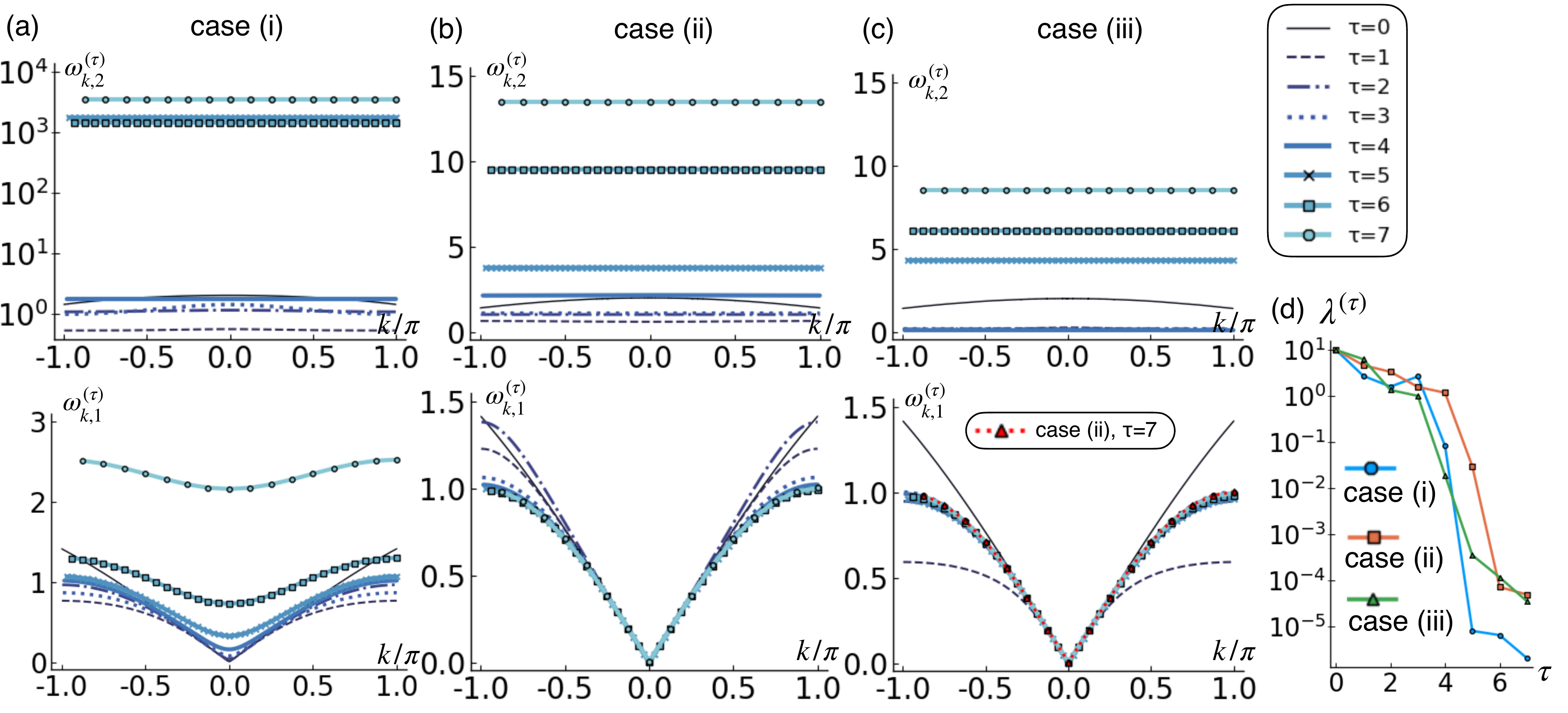}
    \caption{{\bf ER of Finite Temperature 1d Free Boson} (a)-(c) The dispersions of the harmonic modes under $\tau$-steps of RG for cases (i)-(iii) corresponding to gapped, critical, and critical with irrelevant perturbation systems. Note that in (c), we also plot the case (ii) dispersion at the RG step $\tau=7$ for comparison. 
    (d)The quantum perturbation strength $\lambda^{(\tau)}$ at $\tau$-th RG step.
    }
    \label{fig:bosonresult}
\end{figure*}

\textit{ER of Bosonic Gaussian TFD.---}Our next application is the ER of the Bosonic Gaussian TFD in 1d with $N$ sites. 
The application of ER on the ground state and low-energy excitations has been studied in Ref.~\cite{Evenbly_2010}.
We consider the following model,
\begin{align}\label{eqn:bosonH}
    H = \sum_{x=1}^{N} \frac{\lambda^2}{2}\hat{p_x}^2+\hat{V}~,
\end{align}
where 
\begin{align}
   \hat{V} &= \frac{1}{2}\sum_{x=1}^{N}[ \alpha_0^2\hat{q}_x^2 + \alpha_1^2 (\hat{q}_x-\hat{q}_{x+1})^2  \notag \\
    &  + \alpha_2^2 (-\hat{q}_x^2+(\hat{q}_x-\hat{q}_{x+1})^2+\frac{1}{4}(\hat{q}_x+\hat{q}_{x+2})^2) ] ~.
\end{align}

Due to translation invariance and periodic boundary conditions, the modes can be decoupled using Fourier transformation, giving us $\hat{H}=\frac{1}{N}\sum_k \frac{\lambda^2}{2}|\hat{p}_k|^2+ \frac{\omega_k^2}{2}|\hat{q}_k|^2$, where $\omega_k^2=\alpha_0^2+4\alpha_1^2\sin^2(k/2)+4\alpha_2^2\sin^4(k/2)$.
Since the momentum space RG can be carried out exactly, which we summarize in Appendix~\ref{app:momentumRG}, 
we can therefore compare the ER result with the momentum space RG result.

We analyze the ER results with three representative cases: (i) the gapped case $(\alpha_0, \alpha_1, \alpha_2)=(0.1, 1, 0)$, (ii) the ``critical" case $(\alpha_0, \alpha_1, \alpha_2)=(5\! \times\! 10^{-5}, 1, 0)$ and (iii) irrelevant perturbation: $(\alpha_0, \alpha_1, \alpha_2)=(5 \! \times \!10^{-5}, 1, 0.1)$, and show that they are consistent with the momentum space RG results.
In all three cases, we consider $N=2^{12}$, $\beta=1$ and $\lambda=10$.
Note that we explicitly parametrize the $\hat{p}^2$ term with $\lambda$ as the strength of the quantum perturbation.  The correlation function $\la p_{x}\hat{p}_{x^{\prime}}\ra \sim e^{-|x-x^{\prime}|/\xi}$ has a finite correlation length $\xi \propto \beta \lambda$ even when $\omega_k$ is gapless, and thus $\xi$ can be interpreted as a ``quantum correlation length'' which should decrease to zero as the system flows to the classical fixed point.  We note that quantifying quantum correlation lengths generally require more subtle measures such as entanglement negativity \cite{luSingularity2019a,luStructure2020,wuEntanglement2020,luDetecting2020}.

Since the system is quadratic, we can use the covariance matrix to represent the TFD state~\cite{serafiniQuantum2017,holevoEvaluating2001,PhysRevA.66.042327} (see Appendix~\ref{app:cov_matrix of TFD}), which is composed of the correlation functions of $\hat{p}$'s and $\hat{q}$'s.
We also restrict the MERA circuits to be composed of Gaussian gates. 
We can therefore carry out the ER using covariance matrix techniques. (See Appendix~\ref{app:ERbosonGaussianTFG} for the details.)
As illustrated in Fig.~\ref{fig:ER}, we divide $N$ sites into blocks of $2M$ sites.
In general, the disentangling is not perfect, and there will be a small amount of residual entanglement between $G_b$ and the rest of the system.

To compare ER with momentum space RG, we extract the effective reduced density matrix and therefore the effective Hamiltonian. 
Since at each RG step, the system has block-translation invariance, we use Fourier transformation to first decouple the effective density matrix and then obtain the symplectic eigenvalues of the covariance matrix.
Assuming the effective Hamiltonian has the form $\hat{H}_k = \frac{\lambda^2}{2}|\hat{p_k}|^2 + \frac{\omega_k^2}{2}|\hat{q_k}|^2=\lambda\omega_k (\hat{a}_k^{\dagger}\hat{a_k}+\frac{1}{2})$, the symplectic eigenvalues will be $\nu_k=\frac{\exp(\beta \lambda \omega_k)+1}{\exp(\beta \lambda \omega_k)-1}$.
We can therefore infer $\beta \lambda \omega_k = \log(\frac{\nu_k+1}{\nu_k-1})$.
Note that there will be $M$ bands since there are $M$ modes in each block.
Here, analogous to the momentum space RG, we fix $\beta$  and $\frac{\Delta\omega_k}{\Delta k}|_{k=0}$ for the lowest band to be constant  at each RG transformation, and extract the effective $\lambda^{(\tau)}$ and $\omega_{k,a}^{(\tau)}$ where $a=1\hdots M$ at each RG step $\tau$.

We found that $M=2$ can produce results that are consistent with the momentum space RG.
In Fig.~\ref{fig:bosonresult}(a)-(c), we present the results for the three cases (i)-(iii) respectively. 
First, note that in all the three cases, the effective $\lambda$ decreases under the RG (Fig.~\ref{fig:bosonresult}(d)). 
This is indeed desirable that our RG procedure reduces the quantum correlation length, and we can expect that the system will be brought closer and closer to a purely classical system under RG.

For case (i), we can see that the mass term indeed grows exponentially under RG while the band effectively becomes flatter, which is indeed consistent with the conventional momentum space RG result.
In this gapped phase, the system becomes less and less entangled and approaches a product state. 
In particular, one can run the RG until all the correlation lengths become of the order of the lattice spacing (which needs RG steps of the order of $O(\log_2(\xi))$). 
This finite depth ER circuit can therefore be used to construct the TFD state with good precision.

For case (ii), we can see that the small $k$ dispersion stays linear under the RG and is indeed consistent with the momentum space RG as well. 
Note that the small mass term required to protect against the divergence at $k=0$ indeed increases exponentially under RG but is visually negligible in the scale of the figure.
It also appears that, under the RG, the dispersion converges to a fixed dispersion, with a decreasing effective $\lambda$.
In the $\alpha_0 \rightarrow 0$ limit, the system will flow to a classical model with a dispersion $\omega_k \sim |k|$ at small $k$ under ER.
In fact, the entanglement entropy of TFD in case (ii) exhibits a logarithmic scaling in subsystem size, and is described by a Lifshitz critical theory~\cite{heEntanglement2017,mohammadimozaffarEntanglement2017} (see Appendix~\ref{app:EEofgaplessTFD}).

Finally, for case (iii), the dispersion converges under ER to the dispersion of case (ii), which can be seen in Fig.~\ref{fig:bosonresult}(c).
We therefore conclude that the $\alpha_2$ perturbation is irrelevant, which again agrees with the momentum space RG result.

\textit{Discussions.---}In this work, we demonstrate several uses of entanglement renormalization of thermofield double states.  
Such a procedure provides a real space RG scheme for thermal states, and also provides explicit circuits to construct TFDs from simple states.
We have applied the procedure to two nontrivial examples: the toric code TFD and the bosonic Gaussian TFD.
In particular, we have constructed an exact ER circuit which maps the toric code TFD onto a coarse-grained lattice with a renormalized temperature. It is an interesting question whether an exact RG circuit can also be constructed for the 3d and 4d toric code TFD, where the systems at low temperature have classical and topological order, respectively.
For the bosonic Gaussian system, we find the ER procedure can also reproduce the conventional momentum space RG results, such as the irrelevance of the quantum perturbation.  Although we have focused primarily on ER of thermal Gibbs states, our constructions can also be used more generally to generate ER of mixed states.

\begin{acknowledgments}
We thank Tarun Grover, Tsung-Cheng Lu, John McGreevy, and Liujun Zou for valuable discussions and feedbacks.
C.-J.~Lin, Z.~Li\  and T.~H.~Hsieh\ acknowledge support from Perimeter Institute for Theoretical Physics.
This research was supported in part by Perimeter Institute for Theoretical Physics. 
Research at Perimeter Institute is supported in part by the Government of Canada through the Department of Innovation, Science and Economic Development Canada and by the Province of Ontario through the Ministry of Colleges and Universities.
\end{acknowledgments}

%

\clearpage
\onecolumngrid
\appendix

\section{Exact ER circuit of Toric code TFD}\label{app:disentagle_circuit}

\begin{figure}
    \includegraphics[width=0.5\columnwidth]{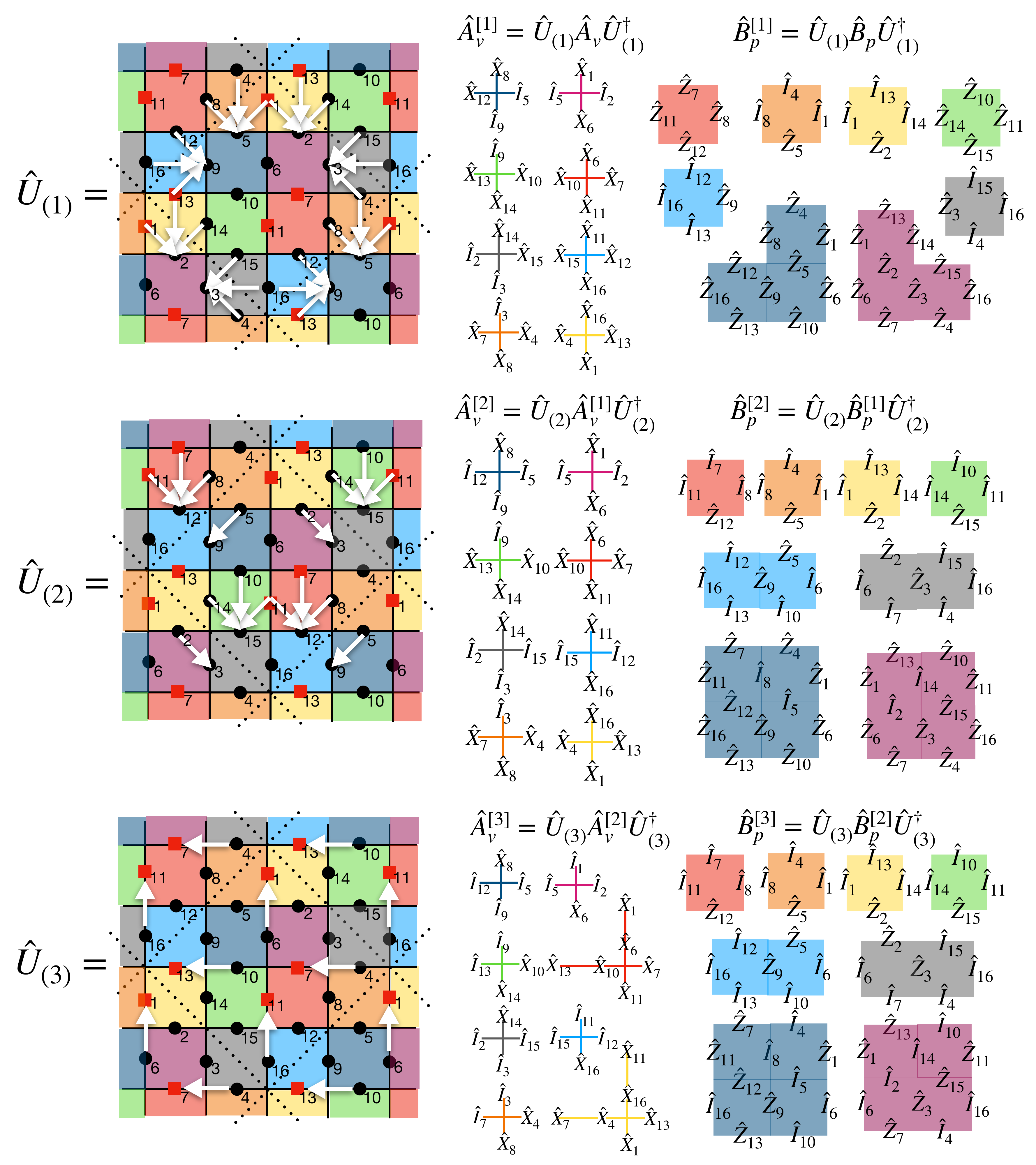}
    \caption{The action of the first three layers of the circuit $\hat{U}_{(1)}$, $\hat{U}_{(2)}$ and $\hat{U}_{(3)}$ operate on the Hamiltonian terms.
    These three circuit operates on the TFD as $\hat{U}_{(i)L}\otimes \hat{U}_{(i)R} |\TFD \ra$ for $i=1,2,3$.
    }
    \label{fig:toricRGcircuit_1}
\end{figure}

In this appendix, we present the detailed construction of the exact ER circuit for the toric code TFD state.
Recall the toric code Hamiltonian is defined as Eq.~(\ref{eqn:toric_code}) in the main text, and we consider its TFD state as in Eq.~(\ref{eqn:generalTFD}) in the main text.
A type of the gate which is heavily used in our exact RG circuit is the CNOT gate, where we denote it as an arrow as shown in Fig.~\ref{fig:toricRGcircuit_1}, where the arrows are pointing from the control qubit to the target qubit. 
It is more convenient to examine how the operators transform under the conjugation of the gates.
Under the CNOT gate, the following operators transform as $IZ \leftrightarrow ZZ$, $ZI \leftrightarrow ZI$, $IX \leftrightarrow IX$ and $XI \leftrightarrow XX$, where the first qubit is the control qubit.

\subsection{Unitaries with no L-R mixing}
The first part of the circuit involves the unitaries without mixing the L and R degrees of freedom, which in fact is the same circuit as constructed in Ref.~\cite{aguadoEntanglement2008}, up to some swaps of the qubits.
In Fig.~\ref{fig:toricRGcircuit_1}, we show the first three unitary transformations $\hat{U}^{(1)}$, $\hat{U}^{(2)}$ and $\hat{U}^{(3)}$ in our RG circuit. 
In the same figure, we also show the transformed Hamiltonian terms $\hat{A}_v^{(n+1)}=\hat{U}_{(n)} \hat{A}_v^{(n)} \hat{U}^{\dagger}_{(n)}$ and $\hat{B}_p^{(n+1)}=\hat{U}_{(n)} \hat{B}_p^{(n)} \hat{U}^{\dagger}_{(n)}$.
In terms of the TFD state, we effectively apply $\hat{U}_{LR}^{(n)}\equiv \hat{U}_L^{(n)} \otimes \hat{U}_R^{(n)}$ consecutively, where $\hat{U}_L^{(n)} = \hat{U}_R^{(n)} = \hat{U}^{(n)}$ for $n=1,2,3$ (not to be confused with the RG steps). 
Defining $|\TFD ^{(n)}\ra = \hat{U}_{LR}^{(n)}|\TFD^{(n-1)} \ra$ and $|\TFD^{(0)} \ra$ as the starting TFD, we have, for $n=1,2,3$,
\begin{align}
    |\TFD ^{(n)}\ra &= \frac{1}{\sqrt{Z_\beta}} \sum_{ \boldsymbol{\sigma}} \hat{U}_L^{(n)} e^{-\frac{\beta}{2}\hat{H}_L^{(n-1)}} |\boldsymbol{\sigma} \ra_{L} \hat{U}_R^{(n)}| \boldsymbol{\sigma}  \ra_{R} \notag \\
    &= \frac{1}{\sqrt{Z_{\beta}}}\sum_{ \boldmath{\sigma} } e^{-\frac{\beta}{2}\hat{H}_L^{(n)}} \hat{U}_L^{(n)} |\boldsymbol{\sigma} \ra_{L} \hat{U}_R^{(n)} |\boldsymbol{\sigma}  \ra_{R} \notag \\
    &= \frac{1}{\sqrt{Z_{\beta}}}\sum_{ \boldsymbol{\sigma} } e^{-\frac{\beta}{2}\hat{H}_L^{(n)} } |\boldsymbol{\sigma} \ra_{L} |\boldsymbol{\sigma}  \ra_{R} \notag~,
\end{align}
where $\hat{H}_L^{(n)}$ is $\hat{H}^{(n)}=-h \sum_v \hat{A}_v ^{(n)} -g \sum_p \hat{B}_p^{(n)}$, composed of the transformed Hamiltonian terms operating on the $L$ degrees of freedom.
Note that in the last line of the equation, we implicitly redefined the basis since the unitaries $\hat{U}_L^{(n)} = \hat{U}_R^{(n)}$.
This part of the circuit can already disentangle the ground state wavefunction of the toric code, as shown in Ref.~\cite{aguadoEntanglement2008}.

\subsection{Circuit generating effective star terms}

\begin{figure}
    \includegraphics[width=0.5\columnwidth]{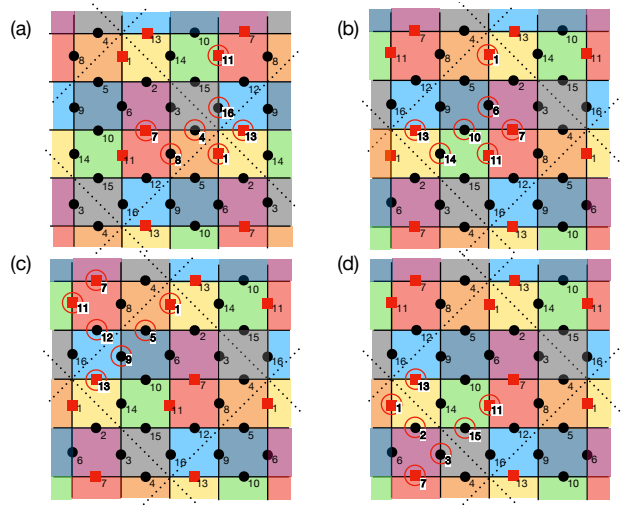}
    \caption{(a)-(d) The highlighted sites in the disentangling procedures. 
    }
    \label{fig:toriccircuit_2}
\end{figure}

Next, we apply the unitaries effectively ``decimating" these degrees of freedom, by disentangling the qubit pairs.
To achieve this, we will need gates which operate across $L$ and $R$ sides.
We first disentangle qubit pairs ($L$ and $R$) on sites $4,8$ and $16$.
As we will see, disentagling these qubit pairs will generate the effective star terms on the coarse-grained lattice.
Note that, since all the star and plaquette terms commute with each other, we can expand
\begin{equation}
    e^{-\frac{\beta}{2} \hat{H}_L^{(3)}}= \prod_{v} (c_{\frac{\beta h}{2}}+s_{\frac{\beta h}{2}} \hat{A}_v^{(3)}) \prod_{p} (c_{\frac{\beta g}{2}}+s_{\frac{\beta g}{2}} \hat{B}_p^{(3)})~,
\end{equation}
where we have abbreviated $c_{x}\equiv \cosh (x)$ and $s_{x}\equiv \sinh (x)$.

Let us first consider the unitary which disentagle the qubit pair on site $8$ highlighted in Fig.~\ref{fig:toriccircuit_2}(a).
We focus on the Hamiltonian terms that involve the site $8$ in the TFD state (the indigo and the orange star terms in $\hat{A}_v^{(3)}$ shown in Fig.~\ref{fig:toricRGcircuit_1}),
\begin{align}
    |\TFD^{(3)} \ra &= \frac{1}{\sqrt{Z_{\beta}}}\sum_{ \boldsymbol{\sigma} } \hat{\rho}_{\text{other}}^{(3)}(c_{\beta h/2} + s_{\beta h/2} \hat{X}_{4L}\hat{X}_{8L})  (c_{\beta h/2} + s_{\beta h/2} \hat{X}_{4L})|\boldsymbol{\sigma} \ra_L |\boldsymbol{\sigma} \ra_R~.
\end{align}
Again, here in the equation, the sites $4$ and $8$ are the specific sites highlighted in Fig.~\ref{fig:toriccircuit_2}(a).
We can see from the composition of the operator, the qubit pair $8$ are entangled with the qubit pair $4$. 
It is also worth noting that, $\hat{\rho}_{\text{other}}^{(3)}$ also contains other operators involving the site $4$ (the yellow star term).
Expanding the operators using $\hat{X}|\sigma \! = \! \pm \ra = \pm  |\pm \ra$ basis, we have 
\begin{align}
    |\TFD^{(3)} \ra &= \frac{1}{\sqrt{Z_{\beta}}}\sum_{ \boldsymbol{\sigma}_{\text{other}} } \hat{\rho}_{\text{other}}^{(3)}|\boldsymbol{\sigma}_{\text{other}}  \ra_L |\boldsymbol{\sigma}_{\text{other}}  \ra_R \notag \\
    &\otimes [|+_L\!\! +_R \ra_4 (e^{\beta h}|+_L\!\! +_R \ra_8+e^{-\beta h}|-_L\!\! -_R \ra_8) 
     + |-_L\!\! -_R \ra_4 (|+_L\!\! +_R \ra_8+|-_L\!\! -_R \ra_8)]
\end{align}
Consider the unitary operator 
\begin{equation}
    \hat{u}^{(4)} = \hat{I} + (|+_L\! \! +_R \ra \la +_L \!\! +_R |)_4 \otimes \hat{v}_8 (\beta h)~,
\end{equation}
where 
\begin{align}\label{eqn:v_i}
    \hat{v}_i (x) &\equiv (\frac{c_x}{\sqrt{c_{2x}}}\! - \! 1)|+_L \!\! +_R \ra \la +_L \!\! +_R |_i -\frac{s_x}{\sqrt{c_{2x}}} |+_L \!\! +_R \ra \la -_L \!\! -_R |_i \notag \\
    &+\frac{s_x}{\sqrt{c_{2x}}} |-_L \!\! -_R \ra \la +_L \!\! +_R |_i +(\frac{c_x}{\sqrt{c_{2x}}}\! - \! 1) |-_L \!\! -_R \ra \la -_L \!\! -_R |_i ~.
\end{align}
It is easy to check that $\hat{u}^{(4)}\hat{u}^{(4)\dagger} = I$ since $\hat{v}_i(x)+\hat{v}_i(x)^{\dagger}+\hat{v}_i(x)\hat{v}_i(x)^{\dagger}=0$.
Furthermore, since the operators in $\hat{\rho}_{\text{other}}^{(3)}$ that touch sites $4$ only involve $\hat{X}_{4L}$'s, it is easy to check $\hat{u}^{(4)}\hat{\rho}_{\text{other}}^{(3)}= \hat{\rho}_{\text{other}}^{(3)}\hat{u}^{(4)}$. 
Operating $\hat{u}^{(4)}$ on the TFD, we have
\begin{align}
    & \hat{u}^{(4)}|\TFD^{(3)} \ra = \frac{1}{\sqrt{Z_{\beta}}}\sum_{ \boldsymbol{\sigma}_{\text{other}} } \hat{\rho}_{\text{other}}^{(3)}|\boldsymbol{\sigma}_{\text{other}}  \ra_L |\boldsymbol{\sigma}_{\text{other}}  \ra_R 
     \otimes(\sqrt{c_{2\beta h}}|+_L \!\! +_R\ra + |-_L \!\! -_R\ra)_4(|+_L \!\! +_R\ra + |-_L \!\! -_R\ra)_8~,
\end{align}
where we see the qubit pair $8L$ and $8R$ are indeed disentangled from the rest of the system.
By defining $\beta^{\prime}h^{\prime}$ such that
\begin{align}\label{eqn:toricRGeq_1}
    c_{\beta^{\prime}h^{\prime}/2}=\frac{1}{2}(1+\frac{1}{\sqrt{c_{2\beta h}}})~, ~~~ s_{\beta^{\prime}h^{\prime}/2}=\frac{1}{2}(1-\frac{1}{\sqrt{c_{2\beta h}}})~,
\end{align}
we can rewrite 
\begin{align}
    & (\sqrt{c_{2\beta h}}|+_L \!\! +_R\ra + |-_L \!\! -_R\ra)_4 =\sum_{\sigma}\sqrt{c_{2\beta h}}(c_{\beta^{\prime}h^{\prime}/2}+s_{\beta^{\prime}h^{\prime}/2}\hat{X}_{4L})|\sigma \ra_{4L}|\sigma \ra_{4R}~.
\end{align}
We therefore construct the unitary $\hat{U}^{(4)} \equiv \bigotimes \hat{u}^{(4)}$, where the tensor product goes over all the counterparts, so that is disentangles all the pairs on sites $8$, giving us $|\TFD ^{(4)} \ra = \hat{U}^{(4)} |\TFD ^{(3)} \ra$.

We proceed to disentangle qubit pairs $16$, by considering a specific site $16$ as highlighted in Fig.~\ref{fig:toriccircuit_2}(a). 
Again, we consider the operators involving the site $16$ (the yellow and blue star terms in $\hat{A}_v^{(3)}$). 
Abbreviating $\hat{X}_{(i_1...i_n)L} \equiv \hat{X}_{i_1L} ... \hat{X}_{i_nL}$,
we write
\begin{align}
    & |\TFD^{(4)} \ra = \frac{1}{\sqrt{Z_{\beta}}}\sum_{\boldsymbol{\sigma} }\hat{\rho}_{\text{other}}^{(4)} (c_{\beta h/2}+s_{\beta h/2}\hat{X}_{16L})(c_{\beta h/2}+s_{\beta h/2}\hat{X}_{(1,4,7,11,13,16)L})|\boldsymbol{\sigma}   \ra_L |\boldsymbol{\sigma}  \ra_R~, \notag \\
\end{align}
where the sites $1,4,7,11,13,16$ are the specific sites highlighted in Fig~\ref{fig:toriccircuit_2}(a).
Again, note that in $\hat{\rho}_{\text{other}}^{(4)}$, there are other terms involving sites $1$, $4$, $7$, $11$ and $13$, but only with $\hat{X}_L$ operators.  

We again expand it in the $\hat{X}$ eigenbasis. 
Denoting $| \pmb{\pm} \ra _{(i_1 ... i_n)} \equiv \sum_{\{\sigma_{i}\}:\sigma_{i_1}...\sigma_{i_n}=\pm 1} |\{\sigma_{i}\} \ra_L |\{\sigma_{i}\} \ra_R$ (configurations of $\sigma$'s multiplied to be $+1$ or $-1$), we rewrite 
\begin{align}
    |\TFD^{(4)} \ra & = \frac{1}{\sqrt{Z_{\beta}}}\sum_{\boldsymbol{\sigma}_{\text{other}}}\hat{\rho}_{\text{other}}^{(4)} |\boldsymbol{\sigma}_{\text{other}}\ra_L |\boldsymbol{\sigma}_{\text{other}}\ra_R \notag \\
    &\otimes[|\pmb{+} \ra_{(1,4,7,11,13)} (e^{\beta h}|+_L\!\! +_R \ra_{16}+e^{-\beta h}|-_L\!\! -_R \ra_{16})+|\pmb{-} \ra_{(1,4,7,11,13)} (|+_L\!\! +_R \ra_{16}+|-_L\!\! -_R \ra_{16})]~.
\end{align}
By abbreviating $|\pmb{+} \ra \la\pmb{+}|\equiv |\pmb{+} \ra_{(1,4,7,11,13)} \la\pmb{+}|_{(1,4,7,11,13)}$ and $\hat{Z}_{(i_1 ... i_n)L}\equiv \hat{Z}_{i_1 L}...\hat{Z}_{i_n L}$, we consider 
\begin{align}
    \hat{u}^{(5)} &= \hat{I} + \hat{v}_{16}(\beta h) \otimes (|\pmb{+} \ra \la\pmb{+}| +  \hat{Z}_{(1,7)L} |\pmb{+} \ra \la\pmb{+}| \hat{Z}_{(1,7)L}+ \hat{Z}_{(1,11)L} |\pmb{+} \ra \la\pmb{+}| \hat{Z}_{(1,11)L} + \hat{Z}_{(1,13)L} |\pmb{+} \ra \la\pmb{+}| \hat{Z}_{(1,13)L}\notag \\
    &+ \hat{Z}_{(7,11)L} |\pmb{+} \ra \la\pmb{+}| \hat{Z}_{(7,11)L} +\hat{Z}_{(7,13)L} |\pmb{+} \ra \la\pmb{+}| \hat{Z}_{(7,13)L} +\hat{Z}_{(11,13)L} |\pmb{+} \ra \la\pmb{+}| \hat{Z}_{(11,13)L} + \hat{Z}_{(1,7,11,13)L} |\pmb{+} \ra \la\pmb{+}| \hat{Z}_{(1,7,11,13)L})~.
\end{align}
One can check that $[\hat{u}^{(5)},\hat{\rho}_{\text{other}}^{(4)}]=0$ and $\hat{u}^{(5)}\hat{u}^{(5)\dagger}=I$.
Operating on $|\TFD^{(4)}\ra$, we have
\begin{align}
    \hat{u}^{(5)}|\TFD^{(4)}\ra = \frac{1}{\sqrt{Z_{\beta}}}\sum_{\boldsymbol{\sigma}_{\text{other}}}\hat{\rho}_{\text{other}}^{(4)} |\boldsymbol{\sigma}_{\text{other}}\ra_L |\boldsymbol{\sigma}_{\text{other}}\ra_R  \otimes (\sqrt{c_{2\beta h}}|\pmb{+}\ra_{(1,4,7,11,13)} + |\pmb{-}\ra_{(1,4,7,11,13)})  \otimes(|+_L\!\! +_R \ra_{16} + |-_L\!\! -_R \ra_{16})~,
\end{align}
where the qubit pair $16$ is disentangled as desired.
Again, we can define $\beta^{\prime}h^{\prime}$ such that 
\begin{align}
    &(\sqrt{c_{2\beta h}}|\pmb{+}\ra_{(1,4,7,11,13)} + |\pmb{-}\ra_{(1,4,7,11,13)})  = \sqrt{c_{2\beta h}} \sum_{\sigma_{(1,4,7,11,13)}}(c_{\beta^{\prime}h^{\prime}/2}+s_{\beta^{\prime}h^{\prime}/2}\hat{X}_{(1,4,7,11,13)L}) |\sigma_{(1,4,7,11,13)}\ra_L |\sigma_{(1,4,7,11,13)}\ra_R~,
\end{align}
where $\sigma_{(1,4,7,11,13)}=(\sigma_1, \sigma_4,\sigma_7,\sigma_{11},\sigma_{13})$ is the spin configuration on the said sites.
We then see that $\hat{U}^{(5)}=\bigotimes \hat{u}^{(5)}$, where the tensor product goes over all the counterparts and therefore can disentangle all the sites of $16$, giving us $|\TFD ^{(5)}\ra= \hat{U}^{(5)}|\TFD ^{(4)}\ra$.

Next, we proceed to disentangle qubit pair $4$, by considering the specific site $4$ highlighted in Fig.~\ref{fig:toriccircuit_2}(a).
Similar to the previous procedure, the TFD state with the relevant operators written out explicitly is 
\begin{align}
    |\TFD^{(5)} \ra 
    &= \frac{1}{\sqrt{Z_{\beta}}}\sum_{\boldsymbol{\sigma}_{\text{other}}}\hat{\rho}_{\text{other}}^{(5)}(c_{\beta^{\prime} h^{\prime}/2}+s_{\beta^{\prime} h^{\prime}/2}\hat{X}_{4L})(c_{\beta^{\prime} h^{\prime}/2}+s_{\beta^{\prime} h^{\prime}/2}\hat{X}_{(1,4,7,11,13)L})|\boldsymbol{\sigma}_{\text{other}}\ra_L |\boldsymbol{\sigma}_{\text{other}}\ra_R \notag \\
    & = \frac{1}{\sqrt{Z_{\beta}}}\sum_{\boldsymbol{\sigma}_{\text{other}}}\hat{\rho}_{\text{other}}^{(5)} |\boldsymbol{\sigma}_{\text{other}}\ra_L |\boldsymbol{\sigma}_{\text{other}}\ra_R \notag \\
    &\otimes[|\pmb{+} \ra_{(1,7,11,13)} (e^{\beta^{\prime} h^{\prime}}|+_L\!\! +_R \ra_{4}+e^{-\beta^{\prime} h^{\prime}}|-_L\!\! -_R \ra_{4})+|\pmb{-} \ra_{(1,7,11,13)} (|+_L\!\! +_R \ra_{4}+|-_L\!\! -_R \ra_{4})]~.
\end{align}

An unitary that disentangles pair $4$ is 
\begin{align}
    \hat{u}^{(6)} &= \hat{I} + \hat{v}_{4}(\beta^{\prime} h^{\prime}) \otimes (|\pmb{+} \ra \la\pmb{+}|  + \hat{Z}_{(1,7)L} |\pmb{+} \ra \la\pmb{+}| \hat{Z}_{(1,7)L}+ \hat{Z}_{(1,11)L} |\pmb{+} \ra \la\pmb{+}| \hat{Z}_{(1,11)L} + \hat{Z}_{(1,13)L} |\pmb{+} \ra \la\pmb{+}| \hat{Z}_{(1,13)L}\notag \\
    &+ \hat{Z}_{(7,11)L} |\pmb{+} \ra \la\pmb{+}| \hat{Z}_{(7,11)L} +\hat{Z}_{(7,13)L} |\pmb{+} \ra \la\pmb{+}| \hat{Z}_{(7,13)L} +\hat{Z}_{(11,13)L} |\pmb{+} \ra \la\pmb{+}| \hat{Z}_{(11,13)L} + \hat{Z}_{(1,7,11,13)L} |\pmb{+} \ra \la\pmb{+}| \hat{Z}_{(1,7,11,13)L})~,
\end{align}
where $|\pmb{+} \ra \la\pmb{+}| = |\pmb{+} \ra_{(1,7,11,13)} \la\pmb{+}|_{(1,7,11,13)}$ in this case.
We therefore have

\begin{align}
    \hat{u}^{(6)}|\TFD^{(5)}\ra & = \frac{1}{\sqrt{Z_{\beta}}}\sum_{\boldsymbol{\sigma}_{\text{other}}}\hat{\rho}_{\text{other}}^{(5)} |\boldsymbol{\sigma}_{\text{other}}\ra_L |\boldsymbol{\sigma}_{\text{other}}\ra_R (\sqrt{c_{2\beta^{\prime} h^{\prime}}}|\pmb{+}\ra_{(1,7,11,13)} + |\pmb{-}\ra_{(1,7,11,13)}) \otimes  (|+_L\!\! +_R \ra_{4} + |-_L\!\! -_R \ra_{4}) \notag \\
    &= \frac{\sqrt{c_{2\beta^{\prime} h^{\prime}}}}{\sqrt{Z_{\beta}}}\sum_{\boldsymbol{\sigma}_{\text{other}}}\hat{\rho}_{\text{other}}^{(5)}   (c_{\beta_{\text{eff}}h_{\text{eff}}/2}+s_{\beta_{\text{eff}}h_{\text{eff}}/2}\hat{X}_{(1,7,11,13)L})|\boldsymbol{\sigma}_{\text{other}}\ra_L |\boldsymbol{\sigma}_{\text{other}}\ra_R \otimes  (|+_L\!\! +_R \ra_{4} + |-_L\!\! -_R \ra_{4})~,
\end{align}

where $\beta_{\text{eff}}h_{\text{eff}}$ is defined as \begin{align}\label{eqn:toricRGeq_2}
    c_{\beta_{\text{eff}} h_{\text{eff}}  /2}=\frac{1}{2}(1+\frac{1}{\sqrt{c_{2 \beta^{\prime} h^{\prime} } } })~,~~~ 
    s_{\beta_{\text{eff}} h_{\text{eff}}  /2}=\frac{1}{2}(1-\frac{1}{\sqrt{c_{2 \beta^{\prime} h^{\prime} } } })
\end{align}
Again, we see that the tensoring all the translation counter part of $\hat{u}^{(6)}$ gives $\hat{U}^{(6)}=\bigotimes \hat{u}^{(6)}$ disentangle all the pairs on sites $4$.
We can also already see that $\hat{X}_{(1,7,11,13)L}$'s are half of the star terms on the coarse-grained lattice.

Disentangling pairs $14$, $6$ and $10$ are done in a similar fashion.
Here we consider the specific highlighted sites in Fig.~\ref{fig:toriccircuit_2}(b), comparing to Fig.~\ref{fig:toriccircuit_2}(a).
It is therefore clear that, we can also generate the desired unitaries by mapping the highlighted sites in Fig.~\ref{fig:toriccircuit_2}(a) to Fig.~\ref{fig:toriccircuit_2}(b).
In particular, to disentangle the pair $14$, we use $\hat{u}^{(7)}$ which is $\hat{u}^{(4)}$ but identifying sites $(8,4)$ in $\hat{u}^{(4)}$ as $(14,10)$; to disentangle the pair $6$, one uses $\hat{u}^{(8)}$ which is $\hat{u}^{(5)}$ but identifying sites $ (11,16,13,4,1,7)$ in $\hat{u}^{(5)}$  as $(1,6,7,10,11,13)$; lastly, to disentangle pairs $10$, we use $\hat{u}^{(9)}$ which is $\hat{u}^{(6)}$ but identifying sites $(11,13,4,1,7)$ in $\hat{u}^{(6)}$ as $(1,7,10,11,13)$.
These circuits result in the star terms $\hat{X}_{(1,7,11,13)L}$ on the highlighted sites in Fig.~\ref{fig:toriccircuit_2}(b) on the coarse-grained lattice with renormalized $\beta_{\text{eff}}h_{\text{eff}}$.

\subsection{Circuit generating effective plaquette terms}
In the last part of the circuit, we aim to decimate the remaining sites $(5,12,9,15,2,3)$, which will generate effective plaquette terms on the coarse-grained lattice.
The construction is in a lot of ways analogous to the previous subsection.

We first disentangle qubit pairs $5$, starting by considering the specific site $5$ as highlighted in Fig.~\ref{fig:toriccircuit_2}(c). 
Writing out the TFD state
\begin{align}
    |\TFD ^{(9)} \ra &= \frac{1}{\sqrt{Z_{\beta}}}\sum_{ \boldsymbol{\sigma} } \hat{\rho}_{\text{other}}^{(9)}(c_{\beta g/2} + s_{\beta g/2} \hat{Z}_{5L}\hat{Z}_{9L}) (c_{\beta g/2} + s_{\beta g/2} \hat{Z}_{5L})|\boldsymbol{\sigma} \ra_L |\boldsymbol{\sigma} \ra_R~.
\end{align}
Here we expand the operator using the basis $\hat{Z}|\pm\ra=\pm |\pm \ra$ and get 
\begin{align}
    |\TFD^{(9)} \ra &= \frac{1}{\sqrt{Z_{\beta}}}\sum_{ \boldsymbol{\sigma}_{\text{other}} } \hat{\rho}_{\text{other}}^{(9)}|\boldsymbol{\sigma}_{\text{other}}  \ra_L |\boldsymbol{\sigma}_{\text{other}}  \ra_R \notag \\
    &\otimes [|+_L\!\! +_R \ra_5 (e^{\beta g}|+_L\!\! +_R \ra_9+e^{-\beta g}|-_L\!\! -_R \ra_9)  + |-_L\!\! -_R \ra_5 (|+_L\!\! +_R \ra_9+|-_L\!\! -_R \ra_9)]~.
\end{align}
Using 
\begin{align}
    \hat{u}^{(10)} = \hat{I} + (|+_L \!\! +_R \ra \la +_L \!\! +_R |)_5 \otimes \hat{v}_9 (\beta g)~,
\end{align}
where $\hat{v}_i (x)$ is given in Eq~(\ref{eqn:v_i}), but with $|\pm \ra$ understood in the $\hat{Z}$ eigenbasis.
Operating $\hat{u}^{(10)}$ to disentangle pair $5$, we have
\begin{align}
    \hat{u}^{(10)}|\TFD^{(9)} \ra &= \frac{1}{\sqrt{Z_{\beta}}}\sum_{ \boldsymbol{\sigma}_{\text{other}} } \hat{\rho}_{\text{other}}^{(9)}|\boldsymbol{\sigma}_{\text{other}}  \ra_L |\boldsymbol{\sigma}_{\text{other}}  \ra_R  \sum_{\sigma}\sqrt{c_{2\beta g}}(c_{\beta^{\prime}g^{\prime}/2}+s_{\beta^{\prime}g^{\prime}/2}\hat{Z}_{9L})|\sigma \ra_{9L}|\sigma \ra_{9R}  \otimes (|+_L \!\! +_R\ra + |-_L \!\! -_R\ra)_5~.
\end{align}

Next, we disentangle pair $12$. 
Similarly, we denote $| \pmb{\pm} \ra _{(i_1 ... i_n)} \equiv \sum_{\{\sigma_{i}\}:\sigma_{i_1}...\sigma_{i_n}=\pm 1} |\{\sigma_{i}\} \ra_L |\{\sigma_{i}\} \ra_R$ (recall we are using $\hat{Z}$ eigenbasis in this subsection), and write 
\begin{align}
    |\TFD^{(10)} \ra 
    & =\frac{1}{\sqrt{Z_{\beta}}}\sum_{\boldsymbol{\sigma} }\hat{\rho}_{\text{other}}^{(10)} (c_{\beta g/2}+s_{\beta g/2}\hat{Z}_{12L})(c_{\beta g/2}+s_{\beta g/2}\hat{Z
    }_{(1,7,9,11,12,13)L})|\boldsymbol{\sigma} \ra_L |\boldsymbol{\sigma}  \ra_R \notag \\
    & = \frac{1}{\sqrt{Z_{\beta}}}\sum_{\boldsymbol{\sigma}_{\text{other}} }\hat{\rho}_{\text{other}}^{(10)} |\boldsymbol{\sigma}_{\text{other}}\ra_L |\boldsymbol{\sigma}_{\text{other}}\ra_R \notag \\
    &\otimes [|\pmb{+} \ra_{(1,7,9,11,13)} (e^{\beta g}|+_L\! +_R \ra_{12}+e^{-\beta g}|-_L\! -_R \ra_{12})+|\pmb{-} \ra_{(1,7,9,11,13)} (|+_L\! +_R \ra_{12}+|-_L\! -_R \ra_{12})]~.
\end{align}
To disentangle pair 12, we use 
\begin{align}
    \hat{u}^{(11)} &= \hat{I} + \hat{v}_{12}(\beta g) \otimes (|\pmb{+} \ra \la\pmb{+}|  + \hat{X}_{(1,7)} |\pmb{+} \ra \la\pmb{+}| \hat{X}_{(1,7)}+ \hat{X}_{(1,11)} |\pmb{+} \ra \la\pmb{+}| \hat{X}_{(1,11)} + \hat{X}_{(1,13)} |\pmb{+} \ra \la\pmb{+}| \hat{X}_{(1,13)} \notag \\
    &+ \hat{X}_{(7,11)} |\pmb{+} \ra \la\pmb{+}| \hat{X}_{(7,11)} +\hat{X}_{(7,13)} |\pmb{+} \ra \la\pmb{+}| \hat{X}_{(7,13)} +\hat{X}_{(11,13)} |\pmb{+} \ra \la\pmb{+}| \hat{X}_{(11,13)} + \hat{X}_{(1,7,11,13)} |\pmb{+} \ra \la\pmb{+}| \hat{X}_{(1,7,11,13)})~,
\end{align}
where $|\pmb{+} \ra \la\pmb{+}| \equiv |\pmb{+} \ra_{(1,7,9,11,13)} \la\pmb{+}|_{(1,7,9,11,13)}$ in this case.
This gives us 
\begin{align}
    \hat{u}^{(11)}|\TFD^{(10)}\ra & = \frac{1}{\sqrt{Z_{\beta}}}\sum_{\boldsymbol{\sigma}_{\text{other}}}\hat{\rho}_{\text{other}}^{(10)} |\boldsymbol{\sigma}_{\text{other}}\ra_L |\boldsymbol{\sigma}_{\text{other}}\ra_R  (\sqrt{c_{2\beta g}}|\pmb{+}\ra_{(1,7,9,11,13)} + |\pmb{-}\ra_{(1,7,9,11,13)}) (|+_L\! +_R \ra_{12} + |-_L\! -_R \ra_{12})~,
\end{align}
with the pair $12$ disentangled. 
We can again express
\begin{align}
    &(\sqrt{c_{2\beta g}}|\pmb{+}\ra_{(1,7,9,11,13)} + |\pmb{-}\ra_{(1,7,9,11,13)})  = \sqrt{c_{2\beta g}} \sum_{\sigma_{(1,7,9,11,13)}}(c_{\beta^{\prime}g^{\prime}/2}+s_{\beta^{\prime}g^{\prime}/2}\hat{Z}_{(1,7,9,11,13)L}) |\sigma_{(1,7,9,11,13)}\ra_L |\sigma_{(1,7,9,11,13)}\ra_R~.
\end{align}

We proceed to disentangle pair $9$.
From
\begin{align}
    |\TFD^{(11)} \ra 
    &= \sum_{\boldsymbol{\sigma}}\hat{\rho}_{\text{other}}^{(11)}(c_{\beta^{\prime} g^{\prime}/2}+s_{\beta^{\prime} g^{\prime}/2}\hat{Z}_{9L})(c_{\beta^{\prime} g^{\prime}/2}+s_{\beta^{\prime} g^{\prime}/2}\hat{Z}_{(1,7,9,11,13)L})|\boldsymbol{\sigma}  \ra_L |\boldsymbol{\sigma}  \ra_R \notag \\
    & = \frac{1}{\sqrt{Z_{\beta}}}\sum_{\boldsymbol{\sigma}_{\text{other}}}\hat{\rho}_{\text{other}}^{(11)} |\boldsymbol{\sigma}_{\text{other}}\ra_L |\boldsymbol{\sigma}_{\text{other}}\ra_R \notag \\
    &[|\pmb{+} \ra_{(1,7,11,13)} (e^{\beta^{\prime} g^{\prime}}|+_L\! +_R \ra_{9}+e^{-\beta^{\prime} g^{\prime}}|-_L\! -_R \ra_{9})+|\pmb{-} \ra_{(1,7,11,13)} (|+_L\! +_R \ra_{9}+|-_L\! -_R \ra_{9})]~,
\end{align}
we use the unitary 
\begin{align}
    \hat{u}^{(12)} &= \hat{I} + \hat{v}_{9}(\beta^{\prime} g^{\prime}) \otimes (|\pmb{+} \ra \la\pmb{+}|  + \hat{X}_{(1,7)} |\pmb{+} \ra \la\pmb{+}| \hat{X}_{(1,7)}+ \hat{X}_{(1,11)} |\pmb{+} \ra \la\pmb{+}| \hat{X}_{(1,11)} + \hat{X}_{(1,13)} |\pmb{+} \ra \la\pmb{+}| \hat{X}_{(1,13)}\notag \\
    &+ \hat{X}_{(7,11)} |\pmb{+} \ra \la\pmb{+}| \hat{X}_{(7,11)} +\hat{X}_{(7,13)} |\pmb{+} \ra \la\pmb{+}| \hat{X}_{(7,13)} +\hat{X}_{(11,13)} |\pmb{+} \ra \la\pmb{+}| \hat{X}_{(11,13)} + \hat{X}_{(1,7,11,13)} |\pmb{+} \ra \la\pmb{+}| \hat{X}_{(1,7,11,13)})~
\end{align}
to disentangle pair $9$, where $|\pmb{+} \ra \la\pmb{+}| \equiv |\pmb{+} \ra_{(1,7,11,13)} \la\pmb{+}|_{(1,7,11,13)}$ in this case.
We have 
\begin{align}
    \hat{u}^{(12)}|\TFD^{(11)}\ra & = \frac{1}{\sqrt{Z_{\beta^{\prime}}}}\sum_{\boldsymbol{\sigma}_{\text{other}}}\hat{\rho}_{\text{other}}^{(11)} |\boldsymbol{\sigma}_{\text{other}}\ra_L |\boldsymbol{\sigma}_{\text{other}}\ra_R  \otimes (\sqrt{c_{2\beta^{\prime} g^{\prime}}}|\pmb{+}\ra_{(1,7,11,13)} + |\pmb{-}\ra_{(1,7,11,13)})  \otimes(|+_L\! +_R \ra_{9} + |-_L\! -_R \ra_{9}) \notag \\
    &= \frac{1}{\sqrt{Z_{\beta^{\prime}}}}\sum_{\boldsymbol{\sigma}_{\text{other}}}\hat{\rho}_{\text{other}}^{(12)} |\boldsymbol{\sigma}_{\text{other}}\ra_L |\boldsymbol{\sigma}_{\text{other}}\ra_R \notag \\
    & \otimes \sqrt{c_{2\beta^{\prime} g^{\prime}}} \sum_{\sigma_{(1,7,11,13)}}(c_{\beta_{\text{eff}}g_{\text{eff}}/2}+s_{\beta_{\text{eff}}g_{\text{eff}}/2}\hat{Z}_{(1,7,11,13)L})   \otimes(|+_L\! +_R \ra_{9} + |-_L\! -_R \ra_{9})~.
\end{align}
We therefore see that half of the plaquette terms are generated on the coarse-grained lattice with renormalized $\beta_{\text{eff}}g_{\text{eff}}$.

Finally, disentangling pairs $2$, $15$ and $3$ as highlighted in Fig.~\ref{fig:toriccircuit_2}(d) is carried out analogously by comparing it to Fig.~\ref{fig:toriccircuit_2}(c). 
To disentangle the pair $2$, we use $\hat{u}^{(13)}$ which is $\hat{u}^{(10)}$ but identifying sites $(5,9) \rightarrow (15,3)$ ; to disentangle the pair $15$, one uses $\hat{u}^{(14)}$ which is $\hat{u}^{(11)}$ but identifying sites $(1,7,9,11,12,13) \rightarrow (11,13,3,1,2,7)$; lastly, to disentangle pairs $3$, we use $\hat{u}^{(15)}$ which is $\hat{u}^{(12)}$ but identifying sites $(1,7,9,11,13) \rightarrow (11,13,3,1,7)$.
These circuit in the end results in the rest of the plaquette terms $\hat{Z}_{(1,7,11,13)}$ on the coarse-grained lattice with renormalized $\beta_{\text{eff}}g_{\text{eff}}$.

\subsection{RG equations}
Here we combine Eqs.~(\ref{eqn:toricRGeq_1}) and Eqs.~(\ref{eqn:toricRGeq_2}) to derive the exact RG equations in the main text.
From Eqs.~(\ref{eqn:toricRGeq_1}), we have 
\begin{equation}
    \tanh\left(\frac{\beta^{\prime} h^{\prime}}{2}\right)=\frac{\cosh^{\frac{1}{4}}(2\beta h)-\cosh^{-\frac{1}{4}}(2\beta h)}{\cosh^{\frac{1}{4}}(2\beta h)+\cosh^{-\frac{1}{4}}(2\beta h)}~,
\end{equation}
or equivalently $\exp(\beta^{\prime} h' /2)=\cosh^{\frac{1}{4}}(2\beta h)$.
We therefore can derive $\tanh(\beta^{\prime} h') = \tanh^2(\beta h)$. 
Similarly, we can get $\tanh(\beta_{\text{eff}} h_{\text{eff}}) = \tanh^2(\beta^{\prime} h')$ from Eqs.~(\ref{eqn:toricRGeq_2}), giving us $\tanh(\beta_{\text{eff}} h_{\text{eff}}) = \tanh^4(\beta h)$ as in the main text. 
The equation for the effective $g$ can be obtained in the same fashion. 

\section{Momentum space RG of the bosonic Gaussian thermal state}\label{app:momentumRG}
In this appendix, we present the results of the momentum space RG for the one dimensional bosonic Gaussian thermal state. 
In particular, we consider harmonic oscillators on a $N$-site chain with the Hamiltonian of Eq.~(\ref{eqn:bosonH}), which we repeat here
\begin{align}
    H = \sum_{x=1}^{N} \frac{\lambda^2}{2}\hat{p_x}^2+\frac{1}{2}\sum_{x,x^\prime}V_{x,x^{\prime}}\hat{q}_x \hat{q}_{x^{\prime}}~.
\end{align}
We consider periodic boundary condition $\hat{p}_{N+1}\equiv \hat{p}_{1}$ and $\hat{q}_{N+1}\equiv \hat{q}_{1}$, and the coupling $V_{x,x^\prime}=V(x-x^{\prime})$ being translation invariant.
Using Fourier transofrmation
\begin{align}
    \hat{q}_x &= \frac{1}{\sqrt{N}}\sum_{k} \hat{q}_k e^{i k x}  \notag \\
    \hat{p}_x &= \frac{1}{\sqrt{N}}\sum_{k} \hat{p}_k e^{-i k x} \notag ~,
\end{align}
where $k = \frac{2\pi n}{N}$ for $n=0~...~N\!-\!1$ and $[\hat{q}_k, \hat{p}_{k^\prime}]=i\delta_{k,k^{\prime}}$,
we have 
\begin{align}
    \hat{H} = \frac{1}{N}\sum_{k}\left(  \frac{\lambda^2}{2}|\hat{p_k}|^2+\frac{1}{2}\omega_k^2|\hat{q}_k|^2 \right) ~,
\end{align}
where $\omega^2_k\equiv\sum_{\ell=1}^{N}V(\ell)e^{-ik\ell}$.
In general, the dispersion can be expanded as $\omega_k^2=\sum_{n=0}^{\infty}c_n |k|^{2n}$. 
In the thermodynamic limit $N \rightarrow \infty$, the Hamiltonian becomes
\begin{align}
    \hat{H} = \int_{-\pi}^{\pi}\frac{dk}{2\pi} \left( \frac{\lambda^2}{2}|\hat{p_k}|^2+\frac{1}{2}\omega_k^2|\hat{q}_k|^2 \right) ~,
\end{align}
The density matrix of the thermal state of the system is $\hat{\rho}(\beta) = \frac{1}{Z(\beta)}e^{-\beta \hat{H}}$ where $Z(\beta)=\tr[e^{-\beta \hat{H}}]$ is the partition function of the system.

To understand the long-distance property of the system, one can coarse grain the system by integrating out the short-distance physics, namely the long-wavelength modes to generate an effective density matrix.
In particular, we consider the RG step by integrating out the modes in $\frac{\pi}{2}<|k|\leq\pi$, which will generate an effective density matrix $\hat{\rho}_{\text{eff}} \equiv \tr_{\frac{\pi}{2}<|k|\leq\pi}[\hat{\rho}]$.
Since all the modes are decoupled, the partial trace can be done exactly, giving us
\begin{align}
    \hat{\rho}_{\text{eff}}=\frac{Z_1(\beta)}{Z(\beta)}\exp \left[-\beta \int_{-\frac{\pi}{2}}^{\frac{\pi}{2}}\frac{dk}{2\pi}\left( \frac{\lambda^2}{2}|\hat{p_k}|^2+\frac{1}{2}\omega_k^2|\hat{q}_k|^2 \right) \right]~,
\end{align}
where $Z_1(\beta)\equiv \tr \exp \left[-\beta \int_{|k|>\frac{\pi}{2}}\frac{dk}{2\pi}\left( \frac{\lambda^2}{2}|\hat{p_k}|^2+\frac{1}{2}\omega_k^2|\hat{q}_k|^2 \right) \right]$.
The next step in RG is to change the length scale by $k^{\prime}=2k$, which gives us 
\begin{align}
    \hat{\rho}_{\text{eff}}=\frac{Z_1(\beta)}{Z(\beta)}\exp \left[-\beta \int_{-\pi}^{\pi}\frac{dk'}{2\pi}\frac{1}{2}\left( \frac{\lambda^2}{2}|\hat{p_{k'}}|^2+\frac{1}{2}\omega_{k'}^2|\hat{q}_{k'}|^2 \right) \right]~,
\end{align}
where $\omega_{k'}^2 = \sum_{n=0}^{\infty}c_n 2^{-2n}|k'|^{2n}$.
Finally, we rescale the mode variables such that the linear term in the dispersion (equivalently $c_1|k'|^2$ term in $\omega_{k'}^2$) is fixed, which requires us to rescale $\hat{q}'_{k'}=\frac{1}{2\sqrt{2}}\hat{q}_{k'}$ and $\hat{p}'_{k'}=2\sqrt{2}\hat{p}_{k'}$ so that $[\hat{q}'_{k}, \hat{p}'_{k'}]=i\delta_{k,k'}$.
We therefore have 
\begin{align}
    \hat{\rho}_{\text{eff}}=\frac{Z_1(\beta)}{Z(\beta)}\exp \left[-\beta \int_{-\pi}^{\pi}\frac{dk}{2\pi}\left( \frac{\lambda_{\text{eff}}^2}{2}|\hat{p}'_{k}|^2+\frac{1}{2}\omega_{\text{eff},k}^2|\hat{q}_{k}'|^2 \right) \right]~,
\end{align}
where $\lambda_{\text{eff}}=2^{-2}\lambda$ and $\omega_{\text{eff},k}^2=\sum_{n=0}^{\infty}c_n 2^{2-2n}|k|^{2n}$.
We therefore see that generally, under the momentum space RG, the quantum perturbation strength (and therefore the quantum correlation length) of the system shrinks, and will resulting in a pure-classical ensemble when the RG step $\tau \rightarrow \infty$. 
For the massless dispersion $c_0=0$, the fixed point ensemble is therefore 

\begin{align}
    \hat{\rho}=\frac{1}{Z(\beta)}\exp \left[-\beta \int_{-\pi}^{\pi}\frac{dk}{2\pi}\left( \frac{1}{2}c_1|k|^2|\hat{q}_{k}|^2 \right) \right]~,
\end{align}
and all the perturbations which contribute to the dispersion $\Delta \omega_k \sim |k|^n$ for $n > 1$ are irrelevant, while the mass term is a relevant perturbation.

\section{Covariance matrix of the Gaussian TFD}\label{app:cov_matrix of TFD}
Given a coupling matrix $V_{x,x^{\prime}}$, the covariance matrix of its thermal state and the purification can be constructed from the $V_{x,x^{\prime}}$ matrix given in Ref.~\cite{holevoEvaluating2001,PhysRevA.66.042327}.
Here, to gain more physical intuitions of the covariance matrix and therefore the correlation function of the system, we directly calculate the correlation functions, assuming the translation invariance of the system.

Assuming, under Fourier transformation, the Hamiltonian can be decoupled into modes
\begin{equation}
    \hat{H}=\frac{1}{N}\sum_k \left(  \frac{\lambda^2}{2}|\hat{p_k}|^2+\frac{1}{2}\omega_k^2|\hat{q}_k|^2 \right) =\frac{1}{N}\sum_{k}\lambda \omega_k (\hat{a}_k^{\dagger}\hat{a}_k+\frac{1}{2})~,
\end{equation}
where 
\begin{align}
    \hat{a}_k &= \frac{1}{\sqrt{2\lambda\omega_k}}(\omega_k \hat{q}_k +i \lambda \hat{p}_{-k})~,~~~~~~~\hat{a}_k^{\dagger} = \frac{1}{\sqrt{2\lambda\omega_k}}(\omega_{k} \hat{q}_{-k} -i \lambda \hat{p}_{k})~,
\end{align}
which are the canonical Boson ladder operators satisfying $[\hat{a}_k, \hat{a}_{k^{\prime}}^{\dagger}]=\delta_{kk^{\prime}}$,
and
\begin{align}
    \hat{q}_k=\sqrt{\frac{\lambda}{2\omega_k}}(\hat{a}_k^{\dagger}+\hat{a}_k)~,~~~~~~~\hat{p}_k=i\sqrt{\frac{\omega_k}{2\lambda}}(\hat{a}_k^{\dagger}-\hat{a}_k)~.
\end{align}

The TFD is therefore 
\begin{equation}
    |\TFD\ra = \bigotimes_k \frac{1}{\sqrt{Z_{\beta,k}}} \sum_{n_k=0}^{\infty} e^{-\frac{\beta}{2} \lambda \omega_k n_k} |n_k\ra_L |n_k\ra_R~,
\end{equation}
where $Z_{\beta,k}\equiv \frac{\exp(-\frac{\beta}{2} \lambda \omega_k)}{1-\exp(-\beta \lambda \omega_k)}$~.
For the later convenience, we define $\la \hat{O} \ra_{\beta} \equiv \tr[e^{-\beta \hat{H}}\hat{O}]/Z_{\beta}$ where $Z_{\beta} \equiv \tr[e^{-\beta \hat{H}}]$ and note that $\bar{n}_k \equiv \la \hat{a}_k^{\dagger} \hat{a}_k  \ra_{\beta} = [\exp(\beta \lambda \omega_k)-1]^{-1}$.

To construct the covariance matrix, we first need $\la\TFD |\hat{q}_{kL}\hat{q}_{k^{\prime}L}|\TFD\ra = \la\TFD| \hat{q}_{kR}\hat{q}_{k^{\prime}R}|\TFD\ra = \la \hat{q}_k \hat{q}_{k^{\prime}} \ra_\beta$, where
\begin{align}
   \la \hat{q}_k \hat{q}_{k^{\prime}} \ra_\beta = \delta_{kk^{\prime}}\frac{\lambda}{2\omega_k}\la (\hat{a}_k^{\dagger})^2+\hat{a}^{\dagger}_k\hat{a}_k +\hat{a}_k\hat{a}_k^{\dagger}+ (\hat{a}_k)^2  \ra_{\beta}=\delta_{kk^{\prime}} \frac{\lambda}{\omega_k}(\bar{n}_k+\frac{1}{2})= \delta_{kk^{\prime}}\frac{\lambda}{2\omega_k}\frac{e^{\beta\lambda\omega_k}+1}{e^{\beta \lambda \omega_k}-1}~,
\end{align}
and similarly $\la\TFD |\hat{p}_{kL}^\dagger\hat{p}_{k^{\prime}L}|\TFD\ra = \la\TFD| \hat{p}_{kR}^\dagger\hat{p}_{k^{\prime}R}|\TFD\ra = \la \hat{p}_k^\dagger \hat{p}_{k^{\prime}} \ra_\beta$, where
\begin{align}
   \la \hat{p}_k \hat{p}_{k^{\prime}} \ra_\beta = -\delta_{kk^{\prime}}\frac{\omega_k}{2\lambda}\la (\hat{a}_k^{\dagger})^2-\hat{a}^{\dagger}_k\hat{a}_k -\hat{a}_k\hat{a}_k^{\dagger} +(\hat{a}_k)^2  \ra_{\beta}=\delta_{kk^{\prime}} \frac{\omega_k}{\lambda}(\bar{n}_k+\frac{1}{2})= \delta_{kk^{\prime}}\frac{\omega_k}{2\lambda}\frac{e^{\beta\lambda\omega_k}+1}{e^{\beta \lambda \omega_k}-1}~.
\end{align}

We also need the correlation functions between the $L$ and $R$ degrees of freedom. 
In particular, $\la \TFD|\hat{q}_{kL} \hat{q}_{k^{\prime}R}|\TFD \ra = \la \TFD|\hat{q}_{kR} \hat{q}_{k^{\prime}L}|\TFD \ra$ and 
\begin{align}
   \la \TFD|\hat{q}_{kL} \hat{q}_{k^{\prime}R}|\TFD\ra &= \delta_{kk^{\prime}}\frac{\lambda}{2\omega_k}\la \TFD| (\hat{a}_{kL}+\hat{a}_{kL}^{\dagger})(\hat{a}_{kR}+\hat{a}_{kR}^{\dagger})|\TFD \ra \notag \\
   &=\delta_{kk^{\prime}}\frac{\lambda}{2\omega_k} \left(\frac{1}{Z_{\beta,k}}\sum_{n,m}e^{-\frac{\beta}{2}\lambda\omega_k (n+m)}n \la m|n\!-\!1\ra^2 + \frac{1}{Z_{\beta,k}}\sum_{n,m}e^{-\frac{\beta}{2}\lambda\omega_k (n+m)}(n+1) \la m|n\!+\!1\ra^2 \right) \notag \\
    &=\delta_{kk^{\prime}}\frac{\lambda}{2\omega_k} \frac{1}{Z_{\beta,k}}\sum_{n}e^{-\beta\lambda\omega_k n}\left( e^{-\frac{\beta}{2}\lambda\omega_k}(n+1)  + e^{\frac{\beta}{2}\lambda\omega_k }n  \right) = \delta_{kk^{\prime}}\frac{\lambda}{\omega_k} \frac{e^{\frac{\beta}{2}\lambda\omega_k}}{e^{\beta\lambda\omega_k}-1}~.
\end{align}
Similarly, $\la \TFD|\hat{p}_{kL} \hat{p}_{k^{\prime}R}|\TFD \ra = \la \TFD|\hat{p}_{kR} \hat{p}_{k^{\prime}L}|\TFD \ra$
\begin{align}
   \la \TFD|\hat{p}_{kL} \hat{p}_{k^{\prime}R}|\TFD\ra &= -\delta_{kk^{\prime}}\frac{\omega_k}{2\lambda}\la \TFD| (\hat{a}_{kL}-\hat{a}_{kL}^{\dagger})(\hat{a}_{kR}-\hat{a}_{kR}^{\dagger})|\TFD \ra  = -\delta_{kk^{\prime}}\frac{\omega_k}{\lambda} \frac{e^{\frac{\beta}{2}\lambda\omega_k}}{e^{\beta\lambda\omega_k}-1}~.
\end{align}

The covariance matrices are the correlation functions in real space.
In particular, $\la \TFD| \hat{r}_{xL(R)} \hat{r}_{xL(R)}| \TFD\ra = \frac{1}{N}\sum_{k}\cos[k(x-x^{\prime})]\la \TFD| \hat{r}_{kL(R)} \hat{r}_{kL(R)}| \TFD\ra$, where $\hat{r}=\hat{q}$ or $\hat{p}$.

It is instructive to examine the long-distance (short-wavelength) behavior of the correlation function for the gapless dispersion. Assuming $\omega_k \sim v_0 |k|$ at small $|k|$, we have 
\begin{equation}
    \la \hat{p}_k \hat{p}_k\ra_{\beta} \approx \frac{2}{\beta \lambda^2}+\frac{1}{6}\beta v_0^2 |k|^2 + O(|k|^4)~.
\end{equation}
This gives us that $\la \hat{p}_x \hat{p}_{x^{\prime}} \ra_{\beta} \sim e^{-|x-x^{\prime}|/\xi}$ where the correlation length $\xi \propto \beta \lambda v_0$.
Note that $\la \hat{p}_x \hat{p}_{x^{\prime}} \ra_{\beta}$ decays exponentially even in the gapless case and such a correlation length can be viewed as a quantum correlation length coming from the quantum perturbation term $\lambda^2\hat{p}^2/2 $.
On the other hand,
\begin{equation}
    \la \hat{q}_k \hat{q}_k\ra_{\beta} \approx \frac{1}{\beta v_0^2 |k|^2}+\frac{1}{6}\beta \lambda^2 + O(|k|^2)~.
\end{equation}
The correlation function $\la \hat{q}_x \hat{q}_{x^{\prime}} \ra_{\beta}$ therefore will diverge in a strictly gapless system in one-dimension. 
To alleviate such a situation, it is customary to consider the dispersion $\omega_k \sim \sqrt{m^2 + v_0^2|k|^2}$ for some small $m$. 
In this case, $\la \hat{q}_x \hat{q}_{x^{\prime}} \ra_{\beta} \sim e^{-|x-x^{\prime}|/\xi}$ with the correlation length $\xi \sim 1/m$. 
One therefore chooses the parameter $m$ such that the correlation length is of the order or greater than the system size $\xi \geq O(N)$ to be the gapless (``critical") system.

\section{Entanglement properties of the Gaussian TFD with the gapless dispersion}\label{app:EEofgaplessTFD}
\begin{figure}
    \includegraphics[width=0.5\columnwidth]{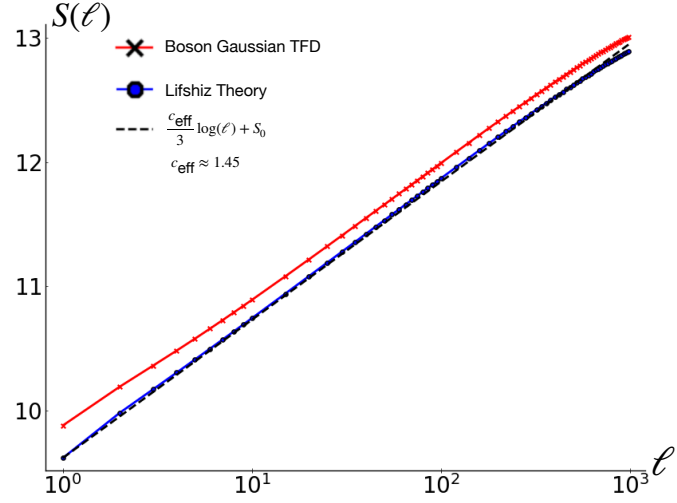}
    \caption{The entanglement entropy scaling with subsystem size $\ell$ for the Boson Gaussian TFD and the Lifshiz theory ($z=2$), with total system size $L=4096$.
    Both states appear to scale logarithmically with subsystem size when $\ell \ll N$.}
    \label{fig:bosonEE}
\end{figure}

It is instructive to examine the entanglement scaling of the Boson TFD for the gapless case (case (ii) in the main text). As we will show below, the logarithmic scaling of the Boson TFD for the gapless case indeed suggests that MERA is a good circuit to capture its entanglement structure.
More specifically, we consider the entanglement entropy of the whole TFD as a function of the subsystem $A$ from site $1$ to $\ell$, where each site contains both $L$ and $R$ degrees of freedom. 
In Fig.~\ref{fig:bosonEE}, we show the scaling of the entanglement entropy $S(\ell)$, which shows a logarithmic behavior. 

In fact, we can understand the origin of this logarithmic scaling. 
The correlation functions of the TFD in the momentum space is given in Appendix~\ref{app:cov_matrix of TFD}.
The nonvanishing correlation function between $L$ and $R$ signifies that the modes are coupled at each $k$ point. 
One can in fact decouple the modes by considering 
\begin{equation}
    \hat{p}^{(\pm)}_k \equiv \frac{1}{\sqrt{2}}(\hat{p}_{kL}\pm\hat{p}_{kR})~, ~~~~~~~ \hat{q}^{(\pm)}_k \equiv \frac{1}{\sqrt{2}}(\hat{q}_{kL}\pm\hat{p}_{qR})~,
\end{equation}
which gives us
\begin{equation}
    \la \hat{p}^{(+)}_k \hat{p}^{(+)}_k \ra = \frac{\omega_k[\cosh(\beta \omega_k /2)-1]}{2\sinh(\beta \omega_k /2)}~,  ~~~~~~ \la \hat{q}^{(+)}_k \hat{q}^{(+)}_k \ra = \frac{[\cosh(\beta \omega_k /2)+1]}{2\omega_k\sinh(\beta \omega_k /2)}~,
\end{equation}
and 
\begin{equation}
    \la \hat{p}^{(-)}_k \hat{p}^{(-)}_k \ra = \frac{\omega_k[\cosh(\beta \omega_k /2)+1]}{2\sinh(\beta \omega_k /2)}~,  ~~~~~~ \la \hat{q}^{(-)}_k \hat{q}^{(-)}_k \ra = \frac{[\cosh(\beta \omega_k /2)-1]}{2\omega_k\sinh(\beta \omega_k /2)}~.
\end{equation}

Comparing with the correlation of the ground state of the Hamiltonian $\hat{H}=\frac{1}{2}|\hat{p}_k|^2 +\frac{1}{2}\Omega_k^2 |\hat{q}_k|^2$ where $\la \hat{p}_k \hat{p}_k \ra = \frac{\Omega_k}{2}$ and $\la \hat{q}_k \hat{q}_k \ra = \frac{1}{2\Omega_k}$, we conclude that the TFD can be viewed as a ground state of the Harmonic oscillators with dispersions $\Omega_k^{(\pm)}=\frac{\omega_k[\cosh(\beta \omega_k/2)\mp 1]}{\sinh(\beta \omega_k/2)}$.

Assume $\omega_k = v_0 |k|$ at small $k$, we find 
\begin{align}
    \Omega_k^{(+)} &\approx \frac{\beta}{4}v_0^2 |k|^2 + O(k^4) \\
    \Omega_k^{(-)} &\approx \frac{4}{\beta}+ \frac{\beta}{12}v_0^2 |k|^2 + O(k^4) ~.
\end{align}
We therefore see that the logarithmic scaling behavior of the entanglement entropy is a result of the $\Omega_k ^{(+)} \sim |k|^2$ dispersion at long-wavelength since $\Omega_k ^{(-)}$ is gapped at low $k$. 
That is, it is described by the $z=2$ Lifshitz theory.
We further confirm this by also calculating the entanglement entropy of the ground state of the Hamiltonian $\hat{H}=\frac{1}{N}\sum_{k}(\frac{1}{2}|\hat{p}_k|^2 +\frac{1}{2}\Omega_k^2 |\hat{q}_k|^2)$ with $\Omega_k^2= m^2 + \sin^4(k/2) $ where $m=10^{-12}$ as shown in Fig.~\ref{fig:bosonEE}.

\section{Details of using covariance matrix for ER in the Bosonic Gaussian problem}\label{app:ERbosonGaussianTFG}

Thanks to the system being quadratic, we use the covariance matrix to represent the TFD state,
\begin{align}
    \Gamma= 
\begin{pmatrix}
    \Gamma_q & O  \\
    O & \Gamma_p  
\end{pmatrix}
\end{align}
where $O$ is a $2N \times 2N$ zero matrix and $\Gamma_{q(p)}$ is a $2N \times 2N$ matrix with entries $[\Gamma_q]_{2i-1,2j-1} = \la \TFD|\hat{q}_{iL} \hat{q}_{jL}  |\TFD \ra$, $[\Gamma_q]_{2i,2j-1} = \la \TFD|\hat{q}_{iR} \hat{q}_{jL}  |\TFD \ra$, $[\Gamma_q]_{2i-1,2j} = \la \TFD|\hat{q}_{iL} \hat{q}_{jR}  |\TFD \ra$ and $[\Gamma_q]_{2i,2j} = \la \TFD|\hat{q}_{iR} \hat{q}_{jR}  |\TFD \ra$ for $i,j=1 ... N$, and similarly for $[\Gamma_p]$ by replacing the $\hat{q}$ operators by $\hat{p}$.

We separate the chain into blocks of $2M$ sites, (therefore $N/2M$ blocks).
We then consider the disentangler operates across the boundary of the blocks, as illustrated in Fig.~\ref{fig:ER} in the main text. 
It is therefore more convenient to further consider the sub-blocks with $M$ sites within each sub-block (therefore $N_b=N/M$ sub-blocks) and each sub-block will have $2M$ harmonic modes from $L$ and $R$.  
Since the system is Gaussian, we expect the Gaussian gates are enough to achieve the desired RG result.
Moreover, we consider the operations that also preserve the $R \leftrightarrow L$ swap symmetry of the TFD. 
In terms of the covariance matrix, we operate
\begin{align}
    \Gamma^{\prime}_p = U_p \Gamma_p U_p^{T}~, ~~~~~~\Gamma^{\prime}_q = U_q \Gamma_q U_q^{T},
\end{align}
where $U_p=(U_q^{-1})^{T}$ and $U_q$ being an invertible matrix with the form
\begin{align}
    U_{q} =  \begin{pmatrix}
    u_q^{(1,1)} & o & o & \hdots & u_q^{(1,N_b)}      \\
        o     & u_q^{(2,2)} & u_q^{(2,3)} &    &  \\
        o & u_q^{(3,2)} & u_q^{(3,3)} &     &  \\
        \vdots       &           &           &\ddots &\\
     u_q^{(N_b,1)} &           &   \hdots        & &u_q^{(N_b,N_b)}
\end{pmatrix}~,
\end{align}
where $u_q^{(n,m)}$ is a $2M \times 2M$ matrix where $n,m$ are the sub-block index and $o$ is a $2M \times 2M$ zero matrix.
To preserve the translational invariance and the periodic boundary condition of the state, we require $u^{(2n+a,2n+b)}=u^{(a,b)}$ where $a,b=0$ or $1$ and $n= 1 \hdots N_b/2$ and we have defined $u^{(N_b+a,N_b+b)}\equiv u^{(a,b)}$ for convenience.
To enforce the $L \leftrightarrow R$ symmetry, we further require the matrix elements $[u^{(a,b)}]_{2i-1,2j-1}=[u^{(a,b)}]_{2i,2j}$ (which are the transformation among $L$ and $R$ degrees of freedom respectively) and $[u^{(a,b)}]_{2i-1,2j}=[u^{(a,b)}]_{2i,2j-1}$ for $a,b = 0,1$ (which mix the $L$ and $R$ degrees of freedom) for $i,j=1 \hdots M$.
There are therefore $8M^2$ parameters for the disentangler.

We then operate the isometry within the blocks, again by Gaussian operations only. 
In terms of covariance matrix, we have 
\begin{align}
    \Gamma_p^{\prime\prime} = W_p \Gamma^{\prime}_p W_p^{T}~,~~~~~~ \Gamma_q^{\prime\prime} = W_q \Gamma^{\prime}_q W_q^{T}
\end{align}
where $W_p=(W_q^{-1})^{T}$ and
\begin{align}
    W_{q} =  \begin{pmatrix}
    w^{(1,1)} & w^{(1,2)} & o & \hdots & & o\\
    w^{(2,1)} & w^{(2,2)} & o &    & &  \\
        o & o & w^{(3,3)} &  w^{(3,4)}   &  \\
        \vdots       &    &w^{(4,3)}       &  w^{(4,4)}         &\ddots &\\
     o &           &   \hdots      &   & &w^{(N,N)}
\end{pmatrix}~.
\end{align}
Again, we require $w^{(2n-a,2n-b)}=w^{(a,b)}$ where $a,b=0$ or $1$ and $n= 1 \hdots N/2$ to enforce block-translation invariance and require  $[w^{(a,b)}]_{2i-1,2j-1}=[w^{(a,b)}]_{2i,2j}$ and $[w^{(a,b)}]_{2i-1,2j}=[w^{(a,b)}]_{2i,2j-1}$ for $a,b = 0,1$ for $i,j=1 \hdots M$ to enforce $L \leftrightarrow R$ symmetry.

We then project out the odd sub-block degree of freedoms, obtaining the course-grained state represented by the density matrix $\Gamma_{q(p),\text{eff}}$ with the matrix element $[\Gamma_{q(p),\text{eff}}]_{i:i+M-1,j:j+M-1}=[\Gamma_{q(p)}^{\prime\prime}]_{2(i-1)+M+1:2(i-1)+2M,2(j-1)+M+1:2(j-1)+2M}$. 
(Note that $[A]_{i_1:j_1, i_2:j_2}$ means the sub-matrix of $A$ with row index from $i_1$ to $j_1$ and column index from $i_2$ to $j_2$.)

The parameters of the gates are obtained by variationally minimizing the entanglement of the first sub-block (the ``garbage" block). 
In terms of covariance matrix, the reduced covariance matrix $\Gamma_{q(p),G_b}$ ($b=1$) is a $2M \times 2M$ matrix with matrix elements $[\Gamma_{q(p),G_{b=1}}]_{i,j}=[\Gamma^{\prime\prime}_{q(p)}]_{i,j}$ for $i,j = 1 \hdots 2M$.
The entanglement entropy is calculated by $S=\sum_{i=1}^{2M}f(\nu_i\!+\!1)-f(\nu_i\!-\!1)$ where $f(x)\equiv [\frac{x}{2}\log \frac{x}{2}]$ and $\nu_i$'s are the symplectic eigenvalues of $\Gamma_{G_b}$. 
Here, following Ref.~\cite{Evenbly_2010}, we minimize $\tr[\Gamma_{G_b}]$ instead as a proxy for minimizing the entanglement entropy.


\end{document}